# Maty's Biography of Abraham De Moivre, Translated, Annotated and Augmented

**David R. Bellhouse and Christian Genest**

*Abstract.* November 27, 2004, marked the 250th anniversary of the death of Abraham De Moivre, best known in statistical circles for his famous large-sample approximation to the binomial distribution, whose generalization is now referred to as the Central Limit Theorem. De Moivre was one of the great pioneers of classical probability theory. He also made seminal contributions in analytic geometry, complex analysis and the theory of annuities. The first biography of De Moivre, on which almost all subsequent ones have since relied, was written in French by Matthew Maty. It was published in 1755 in the *Journal britannique*. The authors provide here, for the first time, a complete translation into English of Maty's biography of De Moivre. New material, much of it taken from modern sources, is given in footnotes, along with numerous annotations designed to provide additional clarity to Maty's biography for contemporary readers.

## INTRODUCTION

Matthew Maty (1718–1776) was born of Huguenot parentage in the city of Utrecht, in Holland. He studied medicine and philosophy at the University of Leiden before immigrating to England in 1740. After a decade in London, he edited for six years the *Journal britannique*, a French-language publication out of the Netherlands that was meant to promote British science and literature throughout continental Europe.

Some time after his arrival in London, Maty became acquainted with Abraham De Moivre. It is possible that their first encounter occurred at Slaughter's Coffee-house, a favorite meeting place of French émigrés that both of them are known to have frequented. In the weeks prior to De Moivre's death, Maty began to interview him in order to write his biography. De Moivre died shortly after giving his reminiscences up to the late 1680s and Maty completed the task using only his own knowledge of the man and De Moivre's published work. The biography, written in French, appeared in the 1755 edition of the *Journal britannique*.

*David R. Bellhouse is Professor, Department of Statistical and Actuarial Sciences, University of Western Ontario, London, Ontario, Canada N6A 5B7 e-mail: bellhouse@stats.uwo.ca. Christian Genest is Professeur titulaire, Département de mathématiques et de statistique, Université Laval, Québec, Canada G1K 7P4 e-mail: genest@mat.ulaval.ca.*

Surviving copies of Maty (1755) are available in only a few locations and are relatively difficult for many to access.[1] More readily available, via *Gallica* on the Internet, is Grandjean de Fouchy's eulogy of De Moivre (Fouchy, 1754). Also written in French, it is based largely on the work of Maty (1755), as Fouchy acknowledges near the end of his text. In fact, his eulogy is for the most part a transcription of excerpts of Maty's biography, with the latter's scientific and personal biases replaced by his own in some places. In spite of appearances, the biography of Maty (1755) predates the eulogy of Fouchy (1754) considerably, since the 1754 volume of the *Histoire de l'Académie royale des sciences* which carried Fouchy's article was actually published in 1759.

---

[1]De Morgan (1846) was possibly the first to refer to Maty (1755) in print. Some 90 years after its publication, Maty's biography of De Moivre was already regarded as obscure by De Morgan, who states: "I can hardly find any notice of this little tract of Dr. Maty." A transcript of Maty (1755) is now available in PDF format on the second author's webpage, at archimede.mat.ulaval.ca/pages/genest.

Since De Moivre's times, concise descriptions of his life and works have been published in several biographical dictionaries, the most recent being Schneider (2004). Maty's article is the major source for almost all of them and remains, to this date, the best and most complete description of this great mathematician's life (Schneider, 2001). On the occasion of the 250th anniversary of De Moivre's death, therefore, it seems fitting to revisit Maty's biography and to provide it in a language that is accessible to a large number of readers. Much additional source material is readily accessible today so that Maty's biographical information has been substantially annotated and augmented. These complements appear in the form of numbered footnotes. Maty's own lettered footnotes to his biography of De Moivre are given as endnotes to the article.

A thorough description and evaluation of De Moivre's mathematical work may be found in Schneider (1968, 2005). Hald (1990) also gives a detailed account of De Moivre's work in probability. Consequently, attention is restricted here mostly to biographical rather than technical detail; the exceptions are when some mathematical commentary enhances Maty's text.

## MEMOIR ON THE LIFE AND WRITINGS OF MR. DE MOIVRE

### By Matthew Maty

I hereby pay tribute to the memory of Mr. De Moivre[2] on behalf of a *Journal britannique* and discharge the duty invested in me through his trust, by

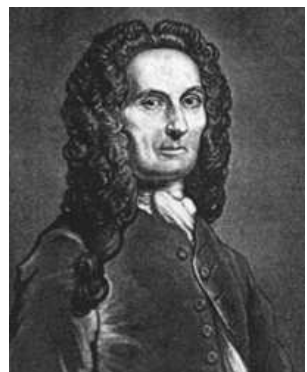

*Abraham De Moivre*
*1667–1754*

publishing what I have been able to gather pertaining to his life and writings. Drawing upon materials that I have collected at pains to myself as well as discoveries that only experts in such matters are competent to appraise, I shall attempt to portray a mathematician who took pride in his own rectitude and who imposed no condition to me other than I speak the language of the truth.

Abraham De Moivre was born at Vitry[3] in Champagne on May 26, 1667. His father was a surgeon and although he was not wealthy, he spared no expense to educate his children.[4] His son was sent to school at an early age, and this son, who retained the fondest memory of his parents throughout his life, recalled with pleasure a letter that he wrote to them on New Year's Day, 1673.

Religious zeal, which was not as keen in this city as in the rest of France, did not preclude Catholic and Protestant families from entrusting their children to the same tutors. Young De Moivre began his studies in Latin with a priest, and after one year, continued his education with the Fathers of the Christian Doctrine. He studied with the latter until the age of eleven and reached third grade. At the time, he was also being taught arithmetic by a tutor of writing. However, one day, when he asked his teacher to clarify an operation on aliquot parts,[5] the latter replied by boxing his ears. This answer was neither to the taste of the young student nor to that of his father; and as the latter was already displeased with the school, he then enrolled his son at the Protestant Academy in Sedan[6]. In the beginning, young De Moivre boarded with the teacher of Greek, whose friendship he won by his devotion to his studies. Although he was among the best students in his class and did not overlook any part of his formal education in the humanities, he found time for studies of a different kind. Assisted only by a fellow student of thirteen years of age, he read a treatise on arithmetic by a certain le Gendre.[7] This is how he learned the rudiments, such as the rule of three, operations on fractions and aliquot parts—the justification for which he had discovered by this time—and even the rule of false position. Whenever his teacher, who was not so keen on arithmetic as he was on Greek, found the table of his pupil forever strewn with calculations, he could not help wondering *what does the little rogue intend to do with those numbers*?

---

[2]Walker (1934) has given an extensive discussion of the spelling of De Moivre's surname. French sources almost invariably refer to the name as Moivre, since the particle "de" would normally be associated with nobility. However, English sources, including De Moivre himself, use De Moivre, de Moivre and Demoivre. De Moivre is used here, since it is the form of his signature on most of his manuscript writings that the authors have been able to see. Schneider (2004) speculates that De Moivre added the particle "de" to his name on arrival in England in order to gain prestige in dealing with English clients, especially among the nobility. This seems doubtful. Among the nearly 1600 Huguenot refugees presenting themselves to the Savoy Church, a French Huguenot church in London, approximately 120 had "de" prefixing their surnames and a further 20 or more had "de la" (Huguenot Society, 1914). If the person was of noble origin, it was noted in the register; for example, Louis de Saint-Delis, Marquis de Heucourt, and Jean de Meslin, Seigneur de Campagny. The vast majority of the 120 entries were not from the nobility.

[3]Modern-day Vitry-le-François, a rebuilding of "Vitry en Perthois," is located in the Department of Marne, about halfway between Paris and Nancy in North-Eastern France.

[4]In a petition for English citizenship (naturalization), Abraham De Moivre states that his parents were named Daniel and Anne (Huguenot Society, 1923). De Moivre had at least one brother, also named Daniel, who was a musician and composer in London; see Lasocki (1989). Schneider (2004) states that this brother became a merchant. However, it was more likely Daniel's son, another Daniel, who was the merchant. The latter Daniel made a business trip to Mexico to buy jewelry in the early 1730s (PRO C104/266 Bundle 38).

After some time with the professor of Greek, Mr. De Moivre pursued his education under the famous humanist Mr. du Rondel,[8] who was in charge of first grade, also known as *Rhetoric*. He remained there until he reached the age of thirteen, that is, until the year 1680, famed for its comet.[9] It had been intended that he enter the Academy under Mr. Bayle after the summer holidays the following year, but his plans were thrown into confusion by the suppression of the Academy and, for lack of a tutor, he was forced to return to Champagne.

The progress he had made in arithmetic was meteoric. Thus his father was advised to find someone to teach him algebra but he had sufficient confidence in his son's ability simply to place in his hands the book by Father Prestet.[10] Unfortunately, the young man found in the introduction to this treatise a preliminary discussion on the nature of our ideas, and since he did not know what an idea was—he had never had the good fortune to hear Mr. Bayle on the subject—he closed the book without ever reading it.

When he was fifteen, he was sent to the Academy in Saumur,[11] where he took his year in *Logic*. His teacher, who instructed him to attend the classes of the Scotsman Duncan,[12] was a poor physicist who had scant esteem for Descartes and who cited no other reason for his contempt than the fact that *he was born before him.*

Such a teacher was obviously unsuitable for such a gifted student. The latter's wish was to be sent to

---

[5]The aliquot parts of a number are the set of proper divisors of the number. For example, the aliquot parts of 12 are 1, 2, 3, 4 and 6.

[6]Sedan is located in the Meuse valley, at the Belgian border, North-East of Paris. The Protestant Academy in Sedan was founded in 1579 at the initiative of Françoise de Bourbon, widow of Henri-Robert de la Marck. The principality of Sedan became part of France in 1642.

[7]François Le Gendre wrote *L'Arithmétique en sa perfection, mise en pratique selon l'usage des financiers, banquiers, & marchands* (Le Gendre, 1657). A description of Le Gendre's work is given in Sanford (1936).

[8]Jacques du Rondel (ca. 1630–1715) and Pierre Bayle (1647–1706) were both teachers at the Huguenot Academy in Sedan. When it was suppressed, du Rondel became a professor of belles lettres at the University of Maestricht, in the Netherlands; his most famous published work is *De Vita et Moribus Epicuri* (On the life and death of Epicurus), which appeared in 1693. As for Bayle, he moved to Rotterdam, where he taught philosophy and history at the École illustre. In 1684 he founded the *Nouvelles de la république des lettres*, the most influential literary and philosophical review of the time; he is most acclaimed for his *Dictionnaire historique et critique* (Bayle, 1696).

[9]This was not Halley's Comet, which appeared in 1682, but rather the "bright comet" discovered by Gottfried Kirch on November 14, 1680, whose position was tracked for several months by astronomers throughout Europe and which Newton used as an illustration of his method of fitting parabolic orbits for comets in the 1687 edition of *Principia Mathematica*.

[10]Jean Prestet (ca. 1648–1690), who taught mathematics in Nantes and Angers, popularized Descartes' principles in his writings. The book referred to here is most likely *Éléments de mathématiques* (Prestet, 1675).

[11]Saumur is located in the Loire valley, some 300 kilometers South-West of Paris, between Tours and Nantes. The Protestant Academy in Saumur was founded in 1589 by Duplesis-Mornay, a friend of King Henri IV. It is noteworthy that Descartes began his life's work there.

[12]Mark Duncan (1570–1640) taught philosophy and Greek at Saumur (Rigg and Bakewell, 2004). The Duncan mentioned here is likely one of his three sons, whom the father had given the names Cérisantis, Sainte-Hélène and Montfort. Among these sons, the most likely candidate is Sainte-Hélène, who "took refuge in London where he died in 1697" (Smiles, 1868, page 508).

Paris, and his indulgent father tried once again to accommodate him. The son, who had finally grasped what was an idea, read almost all of Prestet's book on his own before he left Saumur. In addition, he studied a short treatise by Mr. Huygens on games of chance.[13] Although his comprehension of it was far from complete, he never tired of reading this text and extracted from it ideas that proved useful for the investigations that he undertook thereafter.

Mr. De Moivre arrived in Paris in 1684, and the following year, after studying physics at the Collège de Harcourt,[14] he returned to the family home. He traveled thence to Burgundy to keep company with the son of one of his relatives. Searching among some old books, he found a work on Euclidian geometry by Father Fournier.[15] He read the first few pages eagerly, but on discovering that he was incapable of advancing past the Fifth Proposition, he broke down in tears.[16] When he found him reduced to this state, his relative succeeded in consoling him only after he had promised to explain the proposition to him. Thereafter, he had no trouble finishing all six books. He also read Henrion's *Practical Geometry*,[17] he learned trigonometry and the construction of sine tables, and he studied Rohault's treatises on perspective, mechanics and spherical triangles, all of which had just been published along with some posthumous work.[18]

As Euclid's Books XI and XII seemed too advanced for him, our pupil took advantage of his return to Paris, where he accompanied his father, in order to find a tutor. This person was none other than the famous Ozanam,[19] with whom he studied not only the aforementioned books, but also the rudiments of Theodosius.[20] The aging mathematician was often unequal to the task,[21] but as Mr. De Moivre himself commented: *I dissembled, earmarked the lesson and challenged my teacher to a*

---

[13]Christiaan Huygens (1629–1695) is most famous for his contributions to astronomy. He discovered the true shape of the rings of Saturn and, in 1656, patented the first pendulum clock, which greatly increased the accuracy of time measurement. The short treatise referred to here is *De Ratiociniis in Ludo Aleae* (Huygens, 1657), which is considered to be the first printed work on the calculus of probabilities. This work was part of a larger book by Frans van Schooten, *Exercitationes Mathematicae*. It seems curious that the latter would not be mentioned by Maty, as it contained material that would have been of interest to a developing mathematician.

[14]Among others, the famous French writer Jean Racine was also educated at the Collège de Harcourt beginning in 1658, where he met Molière. This collège was located where Lycée Saint-Louis now stands, near the Sorbonne in Paris.

[15]The Jesuit Georges Fournier (1595–1652) published a Latin version of Euclid's *Elements* (Fournier, 1643).

[16]The fifth proposition in Book I of Euclid's *Elements* states: "In isosceles triangles the angles at the base are equal to one another, and if the equal straight lines be produced further, the angles under the base will be equal to one another." This was the first difficult proposition in Euclid, and since many beginners and the dull-witted stumbled over it, it was often referred to as the *pons asinorum*, or "bridge of fools," in the mid-eighteenth century.

[17]Denis Henrion is the pseudonym for Baron Clément Cyriaque de Mangin (d. 1642). The book referred to here is probably *Quatre livres de la géométrie pratique* (Cyriaque de Mangin, 1620).

[18]This is most likely the *Oeuvres posthumes de Mr. Rohault* (Rohault, 1682).

[19]Jacques Ozanam (1640–1717) was a prolific writer of books on mathematics and is best known today for his work on recreational mathematics. Interestingly, the only published connection between De Moivre and Jacques Ozanam appears after Ozanam's death and it is about chess. In a posthumous edition of Ozanam's *Récréations mathématiques et physiques* (Ozanam, 1725, pages 266–269), there are three solutions to the knight's tour problem, which is to cover all 64 squares of a chess board using a knight's move. There is one solution by Montmort, one by De Moivre and one by Jean-Jacques Mairan who supplied the editor of the *Récréations* with the solutions. At the time, Mairan was directeur de l'Académie royale des sciences, in Paris. In a marginal comment in the book, Mairan says that the solutions were obtained in 1722 (three years after Montmort's death). Twiss (1787, pages 138–139) refers to the problem; he states that of the three solutions, "... it [De Moivre's] is the most regular of any I have seen, and the easiest to be learnt." Here is a diagrammatic representation of De Moivre's solution to the knight's tour problem. The tour starts in the upper right-hand corner of the board.

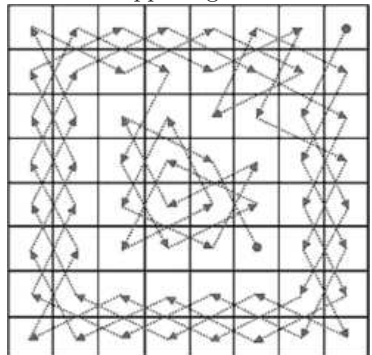

*The graphical solution of the knight's tour problem*

[20]De Moivre was probably studying the *Sphaerics*, written by Theodosius of Bithynia (d. ca. 90 BCE). This was Theodosius's work on the geometry of the sphere, work that provided a mathematical background for astronomy.

[21]Here, Maty is clearly trying to be kind, as Ozanam would have been only 45 years old at the time when he taught De Moivre.

*game of chess.* Little did he say how great was the satisfaction he later derived from discovering on his own what his tutor had been unable to explain!

The tide of religious persecution,[22] which forced many French people to flee to other countries, undoubtedly caused Mr. De Moivre to move to England. At least I have found no other reason why, nor can I say with any precision when he did so, except that he was living there in late 1686,[23] as proved by the following anecdote, which he related to me personally. Once, when he was on his way to pay his respects to the Earl of Devonshire, a distinguished patron of belles-lettres and mathematicians, he saw a man unknown to him leave the Earl's house. The

---

[22] The persecution was the result of King Louis XIV's Edict of Fontainebleau in 1685. This edict revoked the Edict of Nantes of 1598 that had been proclaimed by Louis's grandfather Henri IV. The Edict of Nantes had given substantial rights to French Protestants, known as Huguenots; Henri had been a Protestant and had converted to Catholicism in order to obtain the throne of France. The Edict of Fontainebleau forbade Protestant worship and required all children to be baptized by Catholic priests. Huguenot churches were destroyed and Protestant schools were closed.

[23] In view of the English calendar prior to the calendar reform of 1752, the date given as "in late 1686" could mean as late as March 1687 (new style) since the new year in the old-style calendar began March 25.

The title page of Newton's *Principia* (Newton, 1687) has two printers and two dates given, an earlier date of July 5, 1686, associated with Samuel Pepys and a later date of 1687 associated with Joseph Streater. The book is also printed in two different types which are presumably indicative of the presses of Pepys and Streater. In the anecdote, reference is made to a bound version of the book. It is thus likely to be the finished version of 1687.

There are other sources that lend support to a 1687 date for De Moivre's entry into England. One source is Jacquelot (1712), who described the life and martyrdom of Louis de Marolles, a Huguenot also from Champagne who had been a counsellor to Louis XIV. Marolles had been imprisoned in France by May of 1686. Subsequently, he was sent to the galleys, the penalty for a male Huguenot refusing to convert to Catholicism and trying to leave France. Jacquelot (1712, pages 61–64) mentions that De Moivre was acquainted with Marolles during his confinement. At one point, because Marolles would not abjure, the authorities claimed that he was insane. Marolles responded to the accusation by proposing a mathematics problem that he solved. De Moivre stated that the problem posed was from one given by Ozanam.

There is no further information on De Moivre until the following year. On August 28, 1687, Abraham De Moivre and his brother Daniel presented themselves as Huguenots to be admitted to the Savoy Church in London (Huguenot Society, 1914, page 19). Later that year, on December 16, 1687, the two brothers (their surname spelled phonetically as "de Moavre" in the document), along with several others, were made denizens of England (Cooper, 1862, page 50). Grants of letters patent by the Crown for denization and naturalization (citizenship) were made on various occasions to some Huguenot refugees, usually at a significant cost to the grantee. The designation of denizen allowed some privileges such as ownership of land but fell short of full citizenship.

Abraham De Moivre, but not his brother, did eventually become a full citizen of England. In 1704, Abraham De Moivre was listed on a petition presented to the House of Lords. In the petition, the signatories expressed willingness to serve the Crown in the armed forces. When the names were presented for naturalization in a bill read before the House of Lords, De Moivre's name was not present (Huguenot Society, 1923, page 37 and Royal Commission on Historical Manuscripts, 1910, page 557). Abraham De Moivre's name, among several others, did appear in a naturalization act presented to the Lords in December of 1705. The House of Commons made some amendments and the act received royal assent in March of 1706 (Huguenot Society, 1923, pages 49 and 51 and Royal Commission on Historical Manuscripts, 1912, pages 330–334). Prior to being naturalized, the applicants had to receive the sacrament of Holy Communion in the Church of England. De Moivre, with two of his Huguenot friends, Gideon Nautanie and John Mauries, as well as many other Huguenots, received the sacrament on December 9, 1705, at St. Martin-in-the-Fields church. The three friends each in turn attested to the other two taking communion at the church (Royal Commission on Historical Manuscripts, 1912). Despite the required nominal adherence to the Church of England, Abraham De Moivre probably continued to attend a French Huguenot church in London, in particular West Street Church. His brother Daniel was definitely a member of this parish. Three of Daniel De Moivre's children were baptized at West Street Church: Daniel on January 16, 1707, with his uncle Abraham standing in as godfather; Anne on March 12, 1708, and Elizabeth on June 14, 1709 (Huguenot Society, 1929).

There are two sources that contradict the 1687 arrival in England. Haag and Haag (1846–1859, Volume VII, page 433) state that De Moivre was imprisoned in the Prieuré de Saint-Martin, in Paris, and was not released by the French authorities until April 27, 1688. A reference to source material is given (Arch. Gén. E 3374). An enquiry to the Archives nationales in Paris has resulted in the information that these records have been lost for many years. Agnew (1871, page 84) also states that De Moivre was in the Prieuré de Saint-Martin and was discharged in 1688, although he gives the day as April 15. "Imprisoned," as used by the Haags, is probably too strong a word. The Prieuré de Saint-Martin was a school where Protestant children were sent by the authorities to be indoctrinated into Catholicism. However, the school was not at all successful in converting the children. As Agnew (1871) describes:

> "In the house the boys burnt devotional books, broke images, made an uproar at meal-times, and mixed lumps of lard with fast-day fare. In church they talked or sang where the rubric enjoined silence, moved about from seat to seat, turned their man, who turned out to be Newton, had just left a copy of his *Principia* in the antechamber. Mr. De Moivre was ushered into the same room and took the liberty of opening this book as he waited for the Earl to enter. The illustrations it contained led him to believe that he would have no difficulty reading it. His pride was greatly injured, however, when he realised that he could make neither head nor tail of what he had just read, and that rather than propel him to the forefront of science, as he had anticipated, his studies as a young scholar had merely qualified him for a new development in his career. He rushed out to buy the *Principia*, and as the need to teach mathematics as well as the long walks he was thus forced to take around London left him scarce free time,[24] he would tear out pages from the book and

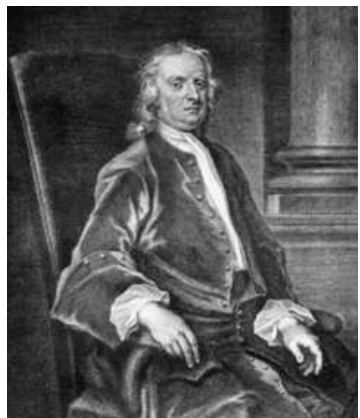

*Sir Isaac Newton*
*1643–1727*

carry them around in his pockets so that he could read them during the intervals between the lessons.[a]

Mr. De Moivre's progress in the science of *infinity* was as swift as it had been in elementary mathematics. He began to establish his reputation. In 1692, he became friends with Mr. Halley[25] and, soon after, Newton[26] himself. The origin and nature of his dealings with the celebrated Mr. Facio[27] speaks of him even more highly. As he was visiting

---

backs on the semi-pagan altar, and stood or sat cross-legged when the congregation knelt."

Agnew notes also that there were many escapes from the priory. This might lead to the explanation that reconciles these sources with the English ones. De Moivre could have escaped from the Prieuré de Saint-Martin a year or more earlier. It was only when the authorities finally gave up or updated their records that they officially discharged him in 1688.

There is some third-hand anecdotal evidence of De Moivre's attitude to the Catholic Church given later in his life. It is attached at the end of a list of Huguenot refugees drawn up by Edward Mangin in 1841. The list is printed in Ewles-Bergeron (1997). The anecdote is:

"I have heard my father say that De Moivre being one day in a Coffee-house in St. Martin's Lane, much frequented by Refugees and other French, overheard a Frenchman say that every good subject ought to be the religion of his King—'Eh quoi donc, Monsieur, si son roi professe la religion du diable, doit-il suivre?"' [Well then, Sir, if his king professes the religion of the devil, should he follow him?]

It is impossible to know whether or not this anecdote, written perhaps 100 years after it originally occurred, is accurate. The beginning of the anecdote, not given here, does contain some inaccuracies. There is reference to De Moivre's daughter rather than to his niece.

[24]Two sources describe De Moivre's work as a teacher or tutor. In a letter to Leibniz dated April 26, 1710, Johann Bernoulli (Leibniz, 1962) referred to De Moivre teaching young boys (he uses the Latin word *adolescentum*) and his state of affairs at the time (*cum fame et miseria*). This assessment is related to complaints that De Moivre had made to Bernoulli nearly two and a half years earlier (Wollenschläger, 1933, page 240). De Moivre said at that time that he taught from morning until night. He was instructing several students during the day and had to walk to where they lived in order to give instruction. He spent a considerable amount of time walking around London.

[25]This is the famous astronomer Edmond Halley (1656–1742), who was also Assistant Secretary to the Royal Society at the time. Cook (1998, page 119) has speculated that De Moivre and Halley first met in Saumur when Halley visited there for about three months in 1681. The meeting of the two at that time is unlikely since, according to his personal recollections given to Maty, De Moivre went to Saumur when he was fifteen years old, which would have been after May, 1682. It is more likely that Halley was introduced to De Moivre through the London Huguenot community, some of whose members were Halley's friends and neighbors in London.

[26]De Moivre became close enough to Newton, probably through many conversations, as to be knowledgeable of the latter's early background and work before they had met. He related these details to John Conduitt, husband of Newton's niece, who was collecting biographical material on Newton a few months after the latter's death. The manuscript of De Moivre's recollections is in the University of Chicago Library.

[27]Nicolas Fatio de Duillier (1664–1753), whose name sometimes appears as Facio, was a Swiss mathematician and close friend of Isaac Newton. He arrived in London the same year as De Moivre, that is, 1687, and was made a fellow of the Royal Society the following year. Fatio was the first to accuse Leibniz of plagiarism in the Newton–Leibniz feud over priority for the discovery of calculus.

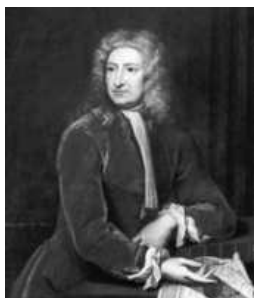

*Edmond Halley*
*1656–1742*

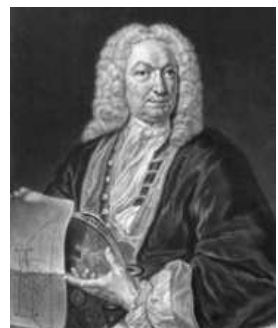

*Johann Bernoulli*
*1667–1748*

a friend named Mr. de Manneville,[28] this mathematician from Geneva once caught him examining a manuscript containing some difficult problems. Mr. Facio asked him banteringly whether he understood them and how he had come upon them. As soon as de Manneville informed him that his teacher was Mr. De Moivre, Mr. Facio wanted the latter to be his teacher, too. He was tutored by him for a month and spread the word that the lessons had been of considerable benefit to him. According to the correspondence between Mr. Leibnitz and Mr. Bernoulli,[b] the same Mr. de Manneville told the latter that for two years, our two mathematicians had sat up whole nights together working on the most abstract topics, among them, the problem of *the most rapidly descending curve.* In the process, I learned that early on, it was Mr. De Moivre's preference to work on difficult problems at night rather than in the day, since they required a great deal of attention; and that, several years later, whenever he felt able to fix his mind on the most complex calculations even during the day, he could not tolerate noise in the house, as the disturbance upset his concentration.[c]

On June 26, 1695, Mr. Halley advised the Royal Society of London that *one Mr. Moivre a French Gentleman has lately discovered to him an improvement*[d] *of the method of fluxions or differentials invented by Mr. Newton with a ready application thereof to rectifying of curve lines, squaring them and their curve surfaces, and finding their centres of gravity, etc.*[29] As a result of Halley's report and no doubt further to Newton's own recommendation, Mr. De Moivre's paper was published in the *Philosophical Transactions* the same year.[30]

It was at some point during this period that Mr. De Moivre devised his general method of raising or

---

[28]Peter de Magneville (d. 1723), phonetically spelled Manneville by Maty, was a Huguenot refugee who eventually lived in London. He appears to have studied with De Moivre and then met Johann Bernoulli on his extensive travels. The Bernoulli–De Moivre correspondence has him at various times in Basel, Frankfurt and Ireland. Bernoulli asked De Moivre to obtain some phosphorus for him and send it to him in Basel; it was Magneville who made the delivery. Magneville died May 8, 1723, while visiting Amsterdam. In his will, besides legacies for his family, he left money to some of his friends including £20 for Abraham De Moivre.

[29]Maty, in his footnote d, says that the quotation in italics was taken from the Registers of the Royal Society. He translated the quotation into French. The version given here is in its original form taken directly from the register or *Journal Book* (pages 307–308) for 1695. That same year, De Moivre helped Halley with one of his papers by providing him with a mathematical result relating to stereographic projections (Halley, 1695). They remained friends for several years. In a 1705 letter (Wollenschläger, 1933, page 198) to the mathematician Johann Bernoulli (1667–1748), De Moivre referred to Halley as "my good and dear friend."

If book ownership is anything to go by, Halley's friendship with De Moivre may have cooled by the late 1720s. According to Osborne (1742), Halley owned copies, at his death in 1742, of most of De Moivre's major books (De Moivre, 1704, 1718 and 1725) as well as a bound copy, separate from the *Philosophical Transactions*, of De Moivre's first publication on probability (De Moivre, 1711). What is missing from the list is De Moivre's major mathematical work, the *Miscellanea Analytica* (De Moivre, 1730) and later editions of *The Doctrine of Chances* and *Annuities upon Lives*; Halley's name does not appear on the subscription list for the *Miscellanea Analytica*.

[30]The paper, which is in the form of a letter, appears as De Moivre (1695). Prior to his introduction to the Royal Society by Halley, De Moivre seems to have been virtually unknown in the mathematical community. When De Moivre (1695) appeared in print, the mathematician John Wallis (1616–1703) wrote on October 24, 1695 to Richard Waller (1646?–1715) at the Royal Society suggesting that some letters written in 1676 from Newton to Henry Oldenburg, then secretary to the Royal Society, should be printed since they "are more to the purpose than that of De Moivre." Prior to this suggestion Wallis commented, "Who this De Moivre is, I know not." See Newton (1959–1977, Volume IV, page 183).

lowering a multinomial $ax + bxx + cx^3 + dx^4$ and so on to any given power.[31] This method entails deriving separately the literary and numerical coefficients of each term in the resulting expression. The literary part consists of the various products of letters whose exponents, represented by their ranks in the alphabet, add up to the power of the desired term. Their form can be deduced from the consideration of the previous terms. As for the numerical factors, they account for reordering. Specifically, the numerical factor associated with each literary product represents the total number of permutations of the letters which compose it.[32] As soon as one sees three or four terms, the regularity of the expression becomes apparent and it can be written down without calculation. Although, as several mathematicians have remarked, this raising or lowering of the multinomial is only a special case of Newton's binomial formula that can be deduced from it, it should be recognized that there is no better way of discovering the pattern according to which each term is formed; any other approach would leave us *wondering about the nature of the terms we face.*[e] The Royal Society, which was apprised of this method in 1697,[33] rewarded its discoverer by making him a member two months later.

The following year, Mr. De Moivre used this theorem to devise a very simple method for reversing a series, that is, for expressing the value of one of the unknowns through a new series consisting of the powers of the other unknown.[34] This method initially seemed less general to Leibnitz than it actually was, so he intended to propose an extension; however, Mr. De Moivre showed that his technique encompassed all the cases that the great mathematician had originally thought to be excluded.[f]

I shall only touch briefly on two or three short writings published in the *Philosophical Transactions*. The first discusses the revolutions of Hippocrates's lune; the second deals with the quadratures of compound curves that have been reduced to simpler ones, and the third describes a particular curve of the third order, similar in several ways to the *foliate*, but different in other respects, just as the ellipse differs from the circle. Although such discoveries might constitute great accomplishments for an ordinary mathematician, they are trifling for a man whose mind is set on loftier achievements.

Mr. De Moivre's career was interrupted by a controversy that was all the more unpleasant since it took a personal turn. In 1703, a Scottish doctor,[35] who has since become famous for a variety of works on theology and medicine, published an essay called *Fluxionum Methodus Inversa*.[36] The subject-matter was new, and the few men capable of making discoveries in this regard were quick to take issue with those who would deprive them of the honor. Mr. Cheyne wronged them by taking the credit for their findings,[37] and although he did not understand their meaning properly, nevertheless attempted to generalize them. Among the plagiarized and disgruntled mathematicians was Mr. De Moivre, who avenged himself the following year by publishing a scathing criticism of Mr. Cheyne's work.[38] The latter's reply carried even more venom in its tail and Mr. De Moivre abandoned the fight. It is clear from Mr. Johann Bernoulli's correspondence and writings how little esteem the great mathematician had for Cheyne's various publications. They also gave rise to a relationship between him and Mr. De Moivre[39] that was as close as can possibly be imagined between two great mathematicians and thus, to some degree, rivals.[40] There was less jealousy and mistrust between Mr. de Varignon and Mr. De Moivre. Indeed, the two of them corresponded with perfect confidence, never quarreled over the priority of their discoveries, and displayed an abiding affection for one another as if they had not both been mathematicians. I would be remiss, should I forget to mention that when Mr. Cheyne gave up mathematics, he showed greater inclination to recognize Mr. De Moivre's merits, and even bought a subscription to one of the latter's major works.[41]

---

[31] The result that Maty is about to describe is often referred to as the "multinomial theorem." It is an extension of Newton's famous "binomial theorem."

[32] To understand what Maty is trying to say here, consider the simple special case in which one wants to determine the coefficient of $x^4$ in the expression $(ax + bx^2 + cx^3 + dx^4 + \cdots)^2$ without expanding the square. Since $1 + 3$ and $2 + 2$ are the only possible decompositions of 4 as the sum of two natural numbers, the "literary" parts of the coefficient would be $ac$ and $bb$, because $a$ and $c$ are respectively the first and third letters of the alphabet, while $b$ is the second letter. (More to point, of course, $a$ and $c$ are the coefficients of $x$ and $x^3$, while $b$ is the coefficient of $x^2$.) As for the associated "numerical" parts of these coefficients, they would be 2 and 1, respectively, because there are two arrangements of factors in the product $ac$, namely $ac$ and $ca$, but only one for $bb$. Consequently, the coefficient of $x^4$ would be $2ac + b^2$, which it is. The technique is valid for infinite polynomials raised to arbitrary integer powers.

[33] The entry in the *Journal Book* for June 16, 1697, reads "Mr. Moivre's paper was read about a method of raising an infinite multinomial to any given power or extracting any given root of the same. He was ordered to have the thanks of the Society and that his paper should be printed." The paper appeared in the *Philosophical Transactions* for that year. On November 30, 1697, the *Journal Book* records that De Moivre and four others that day "were proposed for members, balloted and elected."

[34] For a function expressed, for example, as $y = a_1 x + a_2 x^2 + \cdots$ in a series with no constant term, this would involve writing $x = b_1 y + b_2 y^2 + \cdots$ with $b_1 = a_1^{-1}, b_2 = -a_1^{-3} a_2$, and so on.

[35] George Cheyne (1671–1743) was a Scottish medical doctor who had moved to London. He was one of those physicians interested in applying mathematics to medicine as was his teacher Archibald Pitcairne.

[36] Literally the title translates to "Methods of Inverse Fluxions" or, in modern terms, integral calculus. According to Guicciardini (1989, page 11), the book by Cheyne (1703) was the first attempt in Britain to have a systematic treatment of the calculus.

[37] Maty is a highly sympathetic biographer for his friend De Moivre, and so is taking the "party line" here. Cheyne (1703) made several references in his book to published work. He even asked Newton to look at his manuscript before publishing it. Initially, Newton was favorable to the book and even offered money to Cheyne to get it published. Cheyne de-

---

clined and a misunderstanding ensued. Cheyne wanted Newton to read the work and correct any errors. In the end, it was Joseph Raphson (1648–1715) who did the list of errata published in Cheyne (1703). Newton was apparently offended when the offer of money was declined. D. T. Whiteside, in an introductory section to Newton (1967–1981, Volume VIII), has described in detail the publishing of Cheyne's book and the reaction to it. Whiteside describes the book as "a competent and comprehensive survey of recent developments in the field of 'inverse fluxions' not merely in Britain, at the hands of Newton, David Gregory and John Craige, but also by Leibniz and Johann Bernoulli on the Continent." A contemporary, Humphrey Ditton (1675–1715) also held a balanced view of the dispute and perhaps even mocked the two of them for their fight in the preface to his own book (Ditton, 1706), which was another early work on calculus.

[38] A letter from Varignon to Bernoulli (quoted by Schneider, 1968) suggests that De Moivre's response (De Moivre, 1704) to Cheyne was written at Newton's request. What was really at issue was Newton's failure to publish his work on "quadratures," that is, on finding areas under curves, in 1693. He then let both David Gregory and Edmond Halley see the manuscript, but still did not publish it. What had become clear to Newton was that, in Whiteside's words,

> "... in the ten years since he had penned his revised treatise on quadrature, contemporary techniques for squaring curves had progressed to the point where its propositions were in serious danger of being duplicated..."

In other words, Newton felt threatened by Cheyne's publication.

[39] Prior to the publication of De Moivre (1704), Johann (or Jean, as Maty refers to him in the original French of De Moivre's biography) Bernoulli (1667–1748) did not know who De Moivre was. This is evident in a letter from Bernoulli to Gottfried Leibniz (1646–1716) dated November 29, 1703 (Leibniz, 1962). Bernoulli was passing on to Leibniz information he had received from Cheyne about publications that were in the works in England. He mentioned that a certain De Moivre, whom he knew nothing about, was soon to publish something. The extant correspondence between De Moivre and Bernoulli begins in April of 1704 and continues to 1714; the letters are transcribed in Wollenschläger (1933). From the context of the two earliest letters (De Moivre's first missive to Bernoulli and Bernoulli's reply), it appears that De Moivre initiated the correspondence by sending Bernoulli a copy of his book replying to Cheyne (De Moivre, 1704) along with a letter that made some additional comments on Cheyne's work. Some subsequent letters also discussed Cheyne. The correspondence and friendship continued for a decade, with De Moivre keeping Bernoulli informed of what was happening on the mathematical scene in England. The correspondence came to an end possibly because of the dispute between Leibniz and Newton over priority for the discovery of the calculus. Related to the dispute, Bernoulli had a falling out with Newton.

One of the high points in their relationship occurred in 1712. On October 18, 1712, De Moivre wrote to Bernoulli saying that the mathematicians in England, especially Newton and Halley, were impressed with Bernoulli's latest work. They were going to propose him and his nephew, Nicolaus Bernoulli (1687–1759) for fellowship in the Royal Society. On October 23, 1712, Isaac Newton, in his position as President, proposed Johann Bernoulli for fellowship; he was elected fellow on December 1 (Royal Society *Journal Book*). De Moivre wrote to Bernoulli on December 17 informing him of the election and that it was Newton's idea to postpone the election of Nicolaus. Newton felt that the elder Bernoulli should be elected first, as it would confer on Johann a greater honor. Nicolaus Bernoulli was elected fellow about a year and a half later. Johann Bernoulli wrote back to De Moivre about his election on February 18, 1713, thanking him for the honor and remarking that it was principally De Moivre's efforts that made the election possible. During the time that the election of his uncle to fellowship was underway, Nicolaus Bernoulli was visiting London. De Moivre introduced him to both Newton and Halley. De Moivre and the younger Bernoulli met with Newton three times and dined with him twice.

[40] In his eulogy of De Moivre for the French Academy, which is largely taken from Maty (1755), Fouchy (1754) expresses some doubts about the depth of this relationship as his paraphrase of Maty's biography says: "Some even say that it might have earned him Bernoulli's friendship, had they not been both busy with the same problems, and consequently rivals to a certain extent."

[41] Cheyne subscribed to the *Miscellanea Analytica*.

To divert his friend's mind from these unpleasant events, Dr. Halley encouraged him to turn his attention to astronomy.[42] His advice led to some intriguing findings. In 1705, Mr. De Moivre discovered that *the centripetal force of any planet is directly related to its distance from the centre of the forces and reciprocally related to the product of the diameter of the evolute and the cube of the perpendicular on the tangent.*[43] This theorem, which he stated without proof to Mr. Bernoulli in 1706, was first established by this very knowledgeable professor, who proudly reported it in 1710[44] in a memoir to the Paris Academy of Sciences.[g] Mr. De Moivre pursued further research along those lines. He discovered several very simple properties of conical sections such as, for instance, the fact that *the product of the segments extending from the two foci to any point on an ellipse or a hyperbola is equal to the square of the half-diameter parallel to the tangent.*[45] A simple expression for the principal axes of the ellipse allowed him to solve a number of problems associated with both the general force that maintains planets in their orbits, the points at which the greatest changes in velocity occur, and so on.

In 1706, Mr. De Moivre proposed without proof various formulae for solving, in the manner of Cardan, a large number of equations involving only odd powers of the unknown; these formulae were derived from the consideration of hyperbolic sections. Since the equation of the equilateral hyperbola is the same as that of the circle up to a sign, our scholar applied his formulae to circular arcs, and when Mr. Cotes's treatises appeared posthumously in 1722, Mr. De Moivre was able to use his principles to prove the main theorem. *Suppose that the circumference of a circle with radius* a *is divided into any number* $2\lambda$ *of parts*; *if a line is extended from a point on one of the radii at a distance* $x$ *from the centre of the circle to each of the points*

---

[42]Edmond Halley began his work on comets as early as 1695 (MacPike, 1932). By the next year, based on some calculations he made on the orbits of the comets of 1607 and 1682, he had concluded that the two comets were one and the same. His findings were not published until 1705 at which point he had made calculations on twenty comets and concluded that the comets appearing in 1531, 1607 and 1682 were the same. Now known as Halley's Comet, this comet last appeared in 1986. Halley's work was published in Latin in the *Philosophical Transactions* (Halley, 1705). The article was reprinted in Oxford in pamphlet form, then translated into English and again printed in pamphlet form. On one of the surviving Latin pamphlets, stored at Carnegie Mellon University (Pittsburgh, PA), there is a manuscript letter from De Moivre dated August 25, 1705. The letter is written to a duke (possibly William Cavendish (1641–1707), Duke of Devonshire, who was a fellow of the Royal Society at the time); the opening of the letter reads:

> "The difficulty your Grace has about a passage in Mr. Halley's theory of the comets will I hope be cleared by the following calculation which I would have made sooner and sent your Grace had not I been a little indisposed."

Then follows a number of mathematical calculations related to the velocity of a comet.

[43]In modern notation, the result intuited by De Moivre and later published by Bernoulli (1710) may be described as follows. Suppose that a planet located at point $M$ follows, say, an elliptical orbit whose center of forces is located at focus $F$, as in the picture below. Let $PM$ be the tangent to the curve at $M$, and assume that $FPM$ is a right angle, so that $FP$ is the perpendicular to the tangent. The centripetal force at that point is then proportional to $FM/\{R(FP)^3\}$, where $R$ is the "diameter of the evolute," that is, the radius of curvature at $M$.

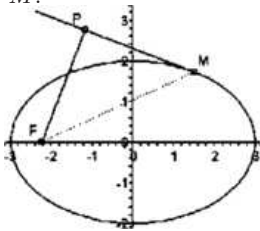

In the special case where the ellipse is circular, $R$ is nothing but the radius of the circle centered at $F$, and $P$ is confounded with $M$, so that the centripetal force is then proportional to $1/(FP)^2$, a classical result of Newtonian mechanics.

[44]Bernoulli (1710) states that he sent a proof of the theorem to De Moivre in a letter dated February 16, 1706. It is not clear, therefore, whether his work was stimulated by De Moivre's conjecture, or whether he knew of the result already. Maty's own wording is equivocal on this point.

[45]The result appears in De Moivre (1717) along with its relationship to centripetal forces. A graphical representation of this fact is as follows:

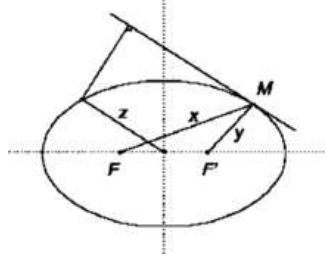

In the graph, $x$ and $y$ are the lengths of the two segments joining the foci to an arbitrary point $M$ on the ellipse, and $z$ is the length of a particular segment going from the origin to the ellipse. Then as stated by Maty, one finds $xy = z^2$, provided that the segment of length $z$ is parallel to the tangent at the point $M$.

*of the division, the product of these lines taken alternately will be equal to the binomial* $a^\lambda + x^\lambda$ *and so on and will give its factors.*[46] Cotes had deduced from this theorem the fluents for an infinite number of fluxions represented by an extremely general expression wherein the quantity had to be restricted, however, to one of the numbers in the sequence 2, 4, 8, 16, and so on. Mr. De Moivre acknowledges somewhere[h] that *his most fervent, abiding wish*— he was always strongly determined—*was that this problem be solved.* It was not long before he found the solution, and he even succeeded in removing the restriction to the powers of 2. The use he made to this end of his discoveries on sections of arcs and angles, as well as particular series—I will return to this matter later—is an analytical marvel.[i] It earned him Mr. Johann Bernoulli's unstinting praise.[k] It was neither of the latter's doing[l] nor, for that matter, was it the fault of Leibnitz, to whom he had been highly recommended and who regarded him as one of England's mathematicians most deserving of esteem,[m] that Mr. De Moivre was not, as he had hoped, appointed to a Chair of Mathematics at some German university—a position that would have rescued him from a form of dependence [on tutoring] that burdened his life more than anyone else's.[47]

The notorious trial surrounding the discovery of these new methods undermined the impartiality that Mr. De Moivre had observed up to that point in the quarrels between the master of German mathematicians and his English counterpart. On April 17, 1712, he was appointed to the Board of Commissioners charged by the Royal Society with examining the old letters in the archives.[48] The names of these commissioners, all of whom have now passed away, are such an integral part of the history of mathematics that they deserve to be mentioned here. They

---

[46]This statement is imprecise and hence somewhat perplexing at first. To clarify its meaning, take $a = 1$ without loss of generality and observe that if $n = 2\lambda$ is an even integer, the roots of $x^n + 1 = 0$ are of the form $\exp\{i\pi(2k-1)/n\}$ for $k = 1, \ldots, n$. Thus

$$x^{2\lambda} + 1 = \prod_{k=1}^{n}[x - \exp\{i\pi(2k-1)/n\}],$$

and the roots divide the unit circle into $2\lambda$ *equal* parts. Maty does not make the latter restriction explicit, however, and he further clouds the issue by speaking of "alternate" products. The thought that he is presumably trying to convey here is that since $n$ is even, the roots can be matched in pairs of the form $e^{i\theta}$ and $e^{-i\theta}$, that is, whose exponents are of alternating sign. Indeed, if $\theta = \pi(2k-1)/n$ for some $k = 1, \ldots, n/2$ and if $l = n - k + 1$, then $\exp\{i\pi(2l-1)/n\} = \exp(-i\theta)$. Furthermore,

$$(x - e^{i\theta})(x - e^{-i\theta}) = x^2 - 2x\cos(\theta) + 1,$$

because $\cos(\theta) = (e^{i\theta} + e^{-i\theta})/2$. This leads to the factorization

$$x^{2\lambda} + 1 = \prod_{k=1}^{n/2}\left[x^2 - 2x\cos\left\{\frac{\pi(2k-1)}{n}\right\} + 1\right],$$

which Cotes (1722) had obtained while working on a number of problems involving logarithmic, trigonometric and hyperbolic functions (Gowing, 1983, page 34). In his *Miscellanea Analytica*, De Moivre (1730) generalized this result, using an equivalent form of what is now known as De Moivre's identity, namely

$$\{\cos(\theta) + i\sin(\theta)\}^n = \cos(n\theta) + i\sin(n\theta).$$

De Moivre used this identity to obtain a factorization formula for any integer $n$ rather than in the special case $n = 2\lambda$. A discussion of De Moivre's work in this area and its relationship to the results of Cotes and Wallis is given in Schneider (1968, pages 237–247). See also Gowing (1983, Chapters 3 and 4).

[47]De Moivre had heard from his friend and former student Magneville that academic positions, chairs of mathematics, were open at two Dutch universities, one at Groningen and the other at Franeker; the latter university closed in 1811. De Moivre wrote to Johann Bernoulli on December 2, 1707, asking his help in obtaining one of these positions, especially the one at Groningen (Wollenshläger, 1933, page 240). Bernoulli, in turn, wrote to Leibniz. Judging by a letter from Leibniz to Bernoulli dated September 6, 1709, nothing had happened by then (see Schneider, 1968, page 207). The only other surviving correspondence on this subject is dated April 26, 1710, at which time Bernoulli asked Leibniz's advice on positions that might be available for De Moivre (Leibniz, 1962).

Maty's footnote m at first glance appears to be a reference to De Moivre's attempt to get an appointment at a "German" university. What is given in Des Maizeaux (1720) is a transcription of a letter from Conti to Newton that is in part praising De Moivre. The part of the letter referring to De Moivre reads [authors' translation] as follows:

> "There is a Frenchman in England, named Mr. de Moivre, whose mathematical knowledge I admire. There are no doubt other skillful people, but who are not totally silent, and from whom you will undoubtedly hear, Sir, & you would oblige me by letting me know."

[48]The Royal Society's *Journal Book* shows that De Moivre, along with two other appointments on the same day (Francis Aston and Brook Taylor), was a late appointment to the Commission. The first six names on Maty's list were appointed March 6, 1712. The Commission reported to the Royal Society on April 24, one week after De Moivre's appointment. He could not have had much impact on the Commission's report.

were MM. Arbuthnot, Hill, Halley, Jones, Machin, Burnet,[n] Robarts, Bonet,[o] De Moivre, Aston, and Taylor. The report that was drawn up and published by these gentlemen with the consent of, and by order of, the Royal Society,[p] is well known.[49] Now that personal and national jealousies are a thing of the past, few people among those who understand the documents used to draft this report fail to agree at least on its main conclusions. [50]

Mr. de Mon[t]mort's *Essay d'analyse sur les jeux de hazard*, published in 1710,[51] almost sparked a similar controversy. Having read this book, Mr. Robart[e]s, who was esteemed for his mathematical erudition at least as much as for his noble extraction,[q] brought the attention of his friend Mr. De Moivre to problems that were more difficult and general than any of those considered therein. The doctrine [theory] of combinations and series, on which the latter had been working diligently for a long time, provided him with the means. He was fueled by his success, and when he eventually became aware of the paths that he and Mr. de Montmort had taken, he was surprised to see how different they were. Hence, he was not afraid to be accused of plagiarizing his work. The Royal Society concurred and ordered that his collection of propositions *De Mensura Sortis*[52] [The Measurement of Chance], which filled a whole issue of the journal, be published in the *Philosophical*

---

[49] The Newton–Leibniz case before the Royal Society is Maty's only mention of De Moivre's activity in the Society. There are other examples of his involvement, however, mostly from the 1730s and beyond. Beginning in 1730, there are eighteen occasions when De Moivre appears as one of the proposers for an individual for fellowship in the Society. Many of his nominees were his students; other nominees can be recognized as having Huguenot origins or émigrés with other national origins; and the balance were continental mathematicians or scientists. These nominations show that De Moivre was active in the Royal Society almost until the end of his life. One year prior to his death, De Moivre was the lead proposer for Robert Symmer (d. 1763) for fellowship. Prior to Newton's death in 1727, De Moivre's nominations probably were done by Newton, as in the case of Johann Bernoulli. There is at least one exception; in 1718 De Moivre proposed Thomas Fantet de Lagny (1660–1734), a French mathematician, for fellowship.

De Moivre was asked by the Royal Society to evaluate the work of at least two individuals, neither of them members of the Society. On one occasion, the Reverend Mr. John Shuttleworth submitted a critique of a treatise on perspective by Lamy (1701). Shuttleworth's claim was that Bernard Lamy (1640–1715) had not taken into account the position of the person's eyes, especially when viewing an object from an angle. In a letter to Shuttleworth (Royal Society), De Moivre refuted the claim and Shuttleworth responded to the Secretary of the Royal Society: "I have sent you Mr. De Moivre's letter. I think he hath not used me candidly in spending so many words upon my letter and saying so little to my treatise. It is, but little encouragement for me to endeavor to perfect the Art of Perspective which L'Amy (tho' a very ingenious author) had not done."

Shuttleworth was never made a fellow of the Royal Society, although he did publish his treatise (Shuttleworth, 1709). De Moivre did look favorably on another publication that he was asked to critique, namely Ludwig Martin Kahle's book on probability (Kahle, 1735). His summary comments at the beginning of his review were:

> "I find that the design of the book is very commendable, it being to shew by several examples that uses that the doctrine of probability may have in common life, and also how the study of it might form the judgement of mankind to a more accurate way of reasoning, than can be derived from common rules of logick."

De Moivre suggested that Kahle be nominated for fellowship in the Royal Society. The name, however, does not appear among the list of fellows.

[50] In view of the documents made available since the mid-nineteenth century, most if not all historians of mathematics would disagree with Maty's conclusion.

[51] Pierre Rémond de Montmort (1678–1719) was a wealthy member of the French aristocracy. His mathematical interests ran from algebra and geometry to probability theory. The *Essay* was actually published in 1708 (Montmort, 1708). According to Rigaud (1841, Volume I, page 256), Montmort sent a copy of the book to the mathematician William Jones (1675–1749) with a covering letter, early in 1709. Montmort may possibly have sent a copy to Francis Robartes (1650–1718) as well. Robartes was a fellow of the Royal Society who was interested in problems in probability. Maty's claim that Robartes encouraged De Moivre to work on problems beyond Montmort's book probably comes from the dedication that De Moivre wrote to Robartes in *De Mensura Sortis* (De Moivre, 1711). De Moivre also mentioned the Robartes connection in a letter that he wrote to Johann Bernoulli in 1712 (Wollenschläger, 1933, page 272) and expanded on it. He said that Robartes had shown him a laborious solution to a probability problem that had involved several cases. The next day, De Moivre found a very simple solution; it appears as Problem 16 in *De Mensura Sortis*. Robartes then posed two more problems and encouraged him to write on probability. During a holiday that he took at a country house, De Moivre finished the manuscript for *De Mensura Sortis* and then submitted it to the Royal Society.

[52] The paper was presented to the Society late in the meeting of June 21, 1711. The original title of the paper was "De Probabilitate Eventum in Ludo Alea" (*Journal Book*, Volume 10, page 305). A translation into English of *De Mensura Sortis* is found in Hald (1984).

*Transactions*.[53] Despite Mr. De Moivre's praise of Mr. de Montmort's work, the latter regarded him as a servile imitator.[54]

He complained to a few friends, and in the second edition of his book, he tried to strip the problems solved in *De Mensura Sortis* of any merit of originality. Yet Mr. De Moivre wanted Mr. de Montmort to be his sole judge. This gave rise to an exchange of letters between them; familiarity and trust appeared to ensue. Our two scholars corresponded with each other about their discoveries on a topic which they treated differently. Mr. de Montmort travelled to London in 1715, *in order*, he wrote to Mr. De Moivre, *to meet with scholars*,[55] *rather than to observe the famous eclipse*.[56] He found in the latter a fellow countryman eager to extend to him all the courtesies of friendship, and so, when he returned to France, he wrote to him and expressed his gratefulness.[57] In 1718, Mr. de Montmort was provided with Mr. De Moivre's second edition, which differed even more significantly than the first edition from anything that he himself had produced. The former died in 1719, without ever repeating his origi-

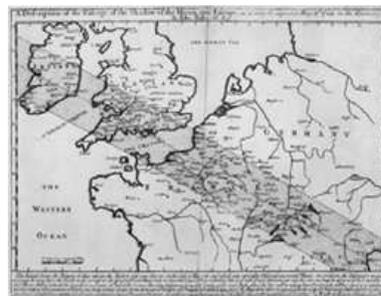

*The map of England and France*

---

[53]De Moivre gave "reprints" of the paper to his friends and close associates. Both Isaac Newton and Edmond Halley had bound versions of the paper in their libraries (Harrison, 1978 and Osborne, 1742). A copy was also sent to Montmort (Montmort, 1713). There is some evidence that De Moivre used *De Mensura Sortis* to advertise or ingratiate himself to potential patrons or clients for his teaching. The Earl of Sunderland received a bound presentation copy from De Moivre with an inscription on the flyleaf (Sunderland, 1881–1883); the current location of this book is unknown.

[54]The praise that De Moivre gave to Montmort was rather muted. In the dedicatory letter to *De Mensura Sortis*, De Moivre stated that, to his knowledge, Huygens was the first to lay down the rules of probability, adding that a French author (unnamed) had recently given several examples of probability calculations that followed these rules. Then De Moivre launched into a description of what was different about his work. In particular, he claimed that his methods were simpler and more general than those of the previous authors. Montmort interpreted these statements as an attack on his work and responded vehemently in the preface to the second edition of *Essay d'analyse* (Montmort, 1713).

[55]In his eulogy of Montmort, de Fontenelle (1719) rather suggests that the main purpose of Montmort's visit to London was to observe the eclipse. Another contemporary source is ambiguous. Halley (1715, page 251) states that he observed the eclipse with several others, naming the Chevalier de Louville as well as Montmort among those in attendance. He notes specifically that Louville was there "purposely to observe the eclipse with us" and took several measurements with the instruments that he had brought with him implying that Montmort was just there to watch the "show."

[56]There was a total solar eclipse over London on May 3, 1715 (April 22, old style). Below is the first-ever eclipse map, produced at the time by Edmond Halley, who also made history by predicting the timing to within four minutes. In the picture, the heavily shaded oval disc represents the umbra or moon's shadow.

[57]Montmort's letters to Brook Taylor (Taylor, 1793) show that, from his point of view, De Moivre and he had patched up their differences. In a letter dated January 2, 1715, Montmort expressed concern over an illness that De Moivre was suffering from. He had also heard that De Moivre was planning a second edition of his work on probability and that it was to be published in English. He referred to the book as "excellent" and expressed his desire for it to be published in Latin so that it would be more widely read. During this time, Montmort sent De Moivre ten theorems on probability that he felt could be included in De Moivre's next edition. By 1716, Montmort was concerned about their scientific relationship and especially the status of his ten theorems. In April of 1716, he wrote to Brook Taylor expressing concern that although he had written De Moivre twice after his visit to England, the latter had not replied. He asked Brook Taylor to look into the matter discretely, saying that he liked De Moivre and thought he was a good man.

One reason why De Moivre may have stopped writing to Montmort is that the latter continued to collaborate with Nicolaus Bernoulli; he would have viewed both as competitors, as they were working on similar problems. The correspondence between Bernoulli and Montmort continued until the latter's death. Further, it was an advantage to De Moivre not to tell the others what he was doing, since he had discovered a new method of solving problems in probability, first using generating functions and then geometrical arguments. Both methods are used in *The Doctrine of Chances* (De Moivre, 1718) without explanation. He would likely have been anxious to keep the method to himself. In confirmation of this, it should be noted that at some point during the time he was preparing *The Doctrine of Chances*, De Moivre wrote a manuscript containing the mathematical background to his methodology. He gave the manuscript to Newton on May 22, 1718, for safekeeping. He explained his position in the preface to *The Doctrine of Chances* (De Moivre, 1718, page ix):

> "Those Demonstrations are omitted purposely to give an occasion for the Reader to exercise his own Ingenuity. In the mean time, I have deposited nal accusations.[58] Nevertheless, a few words[59] in Mr. de Fontenelle's eulogy of Montmort[r] suggest that the French academician's resentment had grown in intensity because he had suppressed it for so long.[60] Despite the praise heaped upon Mr. De Moivre by the illustrious Secretary of the French Academy of Sciences, and the eagerness of the former to show his appreciation through their common friend Mr. de Varignon, Mr. De Moivre nonetheless felt duty-bound to defend himself publicly against the odious suspicion of plagiarism in his *Miscellanea Analytica*,[s] which is the source of my remarks.

Mr. De Moivre's first essay on chance appeared in Latin; the following two editions were published in English and the last one, dated 1738, greatly improved on the earlier ones.[61] The introduction, which lays down the general principles governing calculations on chance, provides the best possible guidelines for anyone wishing to investigate this Logic of likelihoods that Leibnitz called for.[t] Mr. De Moivre describes in the simplest possible terms the underpinnings of the methods presented in his book. The formulae expressing the infinite variety of combinations are sufficient to answer most questions on lotteries and games[62]; a number of other problems,

---

> them with the Royal Society, in order to be published when it shall be thought requisite."

It was shortly after Montmort's death that De Moivre made the manuscript public. Montmort died October 7, 1719, and De Moivre had the manuscript opened at a meeting of the Royal Society on May 5, 1720 (Royal Society, Classified Papers). Some, if not all, of the results in this manuscript appear in De Moivre (1722).

[58]From his letter to Nicolaus Bernoulli dated June 8, 1719, it is clear that Montmort was not only displeased but really infuriated by *The Doctrine of Chances* of 1718, which De Moivre had sent to him as a present. Montmort stated that he wanted nothing more to do with a man like De Moivre who had inserted into his book the results from the second edition of the *Essay* without mentioning either Montmort or Nicolaus Bernoulli (Schneider, 1968, pages 265, 209).

[59]A translation of the relevant words could be as follows: "It is true that he [Montmort] was praised, and is that not sufficient, might one say! But a lord of the manor will not, based on praise alone, release from his obligations a tenant from whom he would expect loyalty and respect for the lands conferred upon him. I [de Fontenelle] speak here as Montmort would have done, without in any way passing judgment as to whether he was in effect the lord."

[60]A possible cause of Montmort's resentment was an engraving that De Moivre included in *The Doctrine of Chances* (De Moivre, 1718). The picture is an allegorical rendering of how De Moivre felt about the importance of his own work when compared to Montmort's. In a dominant posture, the goddess of wisdom is showing the goddess of fortune a diagram by De Moivre that holds the key to his chance calculations, indicating that wisdom now has some hold over chance. The young men in the picture are reading De Moivre's book and have cast aside a chess board, a criticism of Montmort since it was a symbol that appeared in an allegorical picture in the *Essay d'analyse*. The chess board in De Moivre's rendering is not a square one, while in Montmort it is a full board, showing that Montmort's work is incomplete in De Moivre's mind. On the right side of the picture, De Moivre is demonstrating his knowledge of probability to Greek philosophers. The demonstration takes place in the courtyard outside a building that could be interpreted as Aristotle's New Lyceum. A full description of the allegory in the context of the dispute between De Moivre and Montmort is given in Bellhouse (2007b). Montmort was no stranger to allegory, having himself made allegorical allusions to Newton in a sonnet (Taylor, 1793), and would have easily recognized the intent of De Moivre's picture.

[61]There was a posthumous edition published in 1756 (De Moivre, 1756). It was edited by Patrick Murdoch (1710–1774), a mathematician and Church of England clergyman. Earlier, Murdoch had edited a posthumous work of Maclaurin. Confirmation of his editing of De Moivre (1756) is in a letter from Murdoch to Lord Philip Stanhope (1714–1786) dated March 18, 1755 (Centre for Kentish Studies). The letter reads:

> "The Edition which Mr. De Moivre desired me to make of his Chances is now almost printed; and a few things, taken from other parts of his work, are to be subjoined in an Appendix. To which Mr. Stevens, and some other Gentlemen, propose to add some things relating to the same subject; but without naming any author: and he thought if your Lordship was pleased to communicate anything of yours, it would be a favour done the publick. Mr. Scott also tells me, there are in your Lordship's hands two Copy Books containing some propositions on Chances, which De Moivre allowed him to copy. If your Lordship would be pleased to transmit these (to Millar's) with your judgement of them, it might be a great advantage to the Edition."

[62]In *De Mensura Sortis* (De Moivre, 1711), De Moivre made no mention of specific games of chance, generally formulating his problems instead in terms of playing at dice or at bowls. Later, *The Doctrine of Chances* (De Moivre, 1718) contains insightful analyses of particular games played at the time, such as Pharaon and Bassette. The question then arises: did De Moivre gamble? His earlier work uses generic gambling situations as a model; the latter work shows very good knowledge of particular card games. There is no direct evidence of De Moivre gambling at these games. Some circumstantial evidence is that later in life, De Moivre gave advice to gamblers (Le Blanc, 1747, Volume II, page 309). The only other evidence is also circumstantial. In the early 1730s, De Moivre's

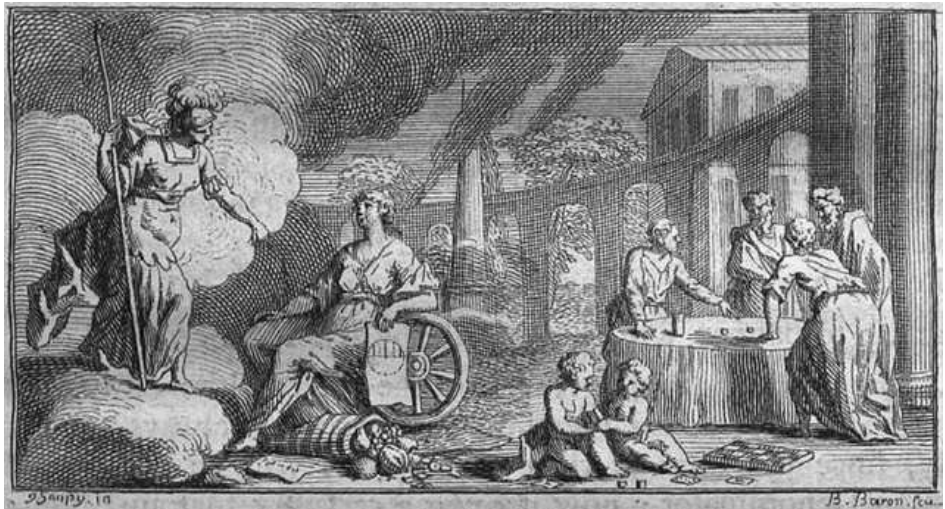

*Allegorical engraving from De Moivre (1718)*

in particular those pertaining to priority and duration of play, can only be solved with the help of series. Those that Mr. De Moivre broaches most often and which he calls *recurrent*[u] are peculiar insofar as each of the terms has a fixed relationship to two or three of those that precede it.[63] As these always break down into a certain number of geometric progressions, their sums can be computed, and one can determine any term or given number of terms thereof. Without the help of approximations, however, the number of operations required would soon become overwhelming. Once more, our scholar's previous discoveries on circular sections supplied him with the means needed for expressing, through the logarithms of sines, the values that he sought. This ingenious application is illustrated in the frontispiece[64] of his book, where a semi-circumference, whose divisions replicate the spokes of a wheel,[65] overlaps a wheel of fortune. If any student, as generous as he is appreciative, were ever to erect a monument to the memory of Mr. De Moivre alongside that of the great Newton, he could have a similar emblem engraved on it, just as a sphere [Maty writes "circle"] inscribed in a cylinder was engraved on the tomb of Archimedes and the logarithmic spiral inscribed on that of the Bernoullis' eldest son.[66]

---

nephew Daniel De Moivre undertook an overseas business venture in which he was required to keep detailed financial records (PRO C104). Over several months of 1731 and 1732, Daniel both won and lost at cards with wins nearly as high as £4 and losses ranging to the same level. Typically his net in any month was about £1 usually on the win side. Like nephew, like uncle?

[63]According to De Moivre (1718, page 133), the terms of a recurrent series are "so related to one another that each of them may have to the same number of preceding terms a certain given relation, always expressible by the same index." The term "recurrent" was only introduced in De Moivre (1722). In modern notation, a series $\sum a_n$ is recurrent if there exist constants $b_1, \ldots, b_k$ such that for all $n > k$, $a_n = b_1 a_{n-1} + b_2 a_{n-2} + \cdots + b_k a_{n-k}$.

[64]Strictly speaking, the picture is not a frontispiece. It appears on page 1 of the book after the title page following by a two-page dedication to Newton and a fourteen-page preface.

[65]The frontispiece is the picture previously mentioned that may have irritated Montmort.

[66]The "semicircumference" mentioned by Maty which should have become De Moivre's epitaph first appears in De Moivre (1722).

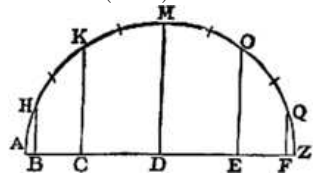

The diagram was actually used, but did not appear, four years earlier in De Moivre (1718). In the first edition of *The Doctrine of Chances*, De Moivre solved the duration of play using recursion methods and then quite abruptly inserted a geometric solution without proof or reference to the diagram. Using modern notation from Hald (1990, page 372), the probability that the duration of play exceeds $n$ games when two players, with probabilities $p$ and $q$ each of winning a game, initially have $b$ stakes each, is given by De Moivre, say for $b$ even, as

$$\sum_{j=1}^{b/2} c_j t_j^{n/2},$$

Speaking of Bernoulli, I am reminded of a problem raised and solved in part in a posthumous treatise of his on the art of conjecture.[x] At issue is *whether it is possible to increase the number of observations of contingent events sufficiently to guarantee with a desired degree of certainty that the number of times they occur will be circumscribed within certain limits*.[67] Mr. Nicolaus Bernoulli, editor of the book written by his uncle but published posthumously, approached the problem from the opposite end by seeking *the probability that would result from a given number of experiments*. But both obtained only partial results, and Nicolaus Bernoulli, who was rather modest about his own accomplishment, considered this problem to be harder than the squaring of the circle.[y] Its solution involves a binomial raised to very high powers and depends on the proportion between the various terms of the binomial raised in this manner. Mr. De Moivre arranged to have a paper on this subject printed for a few friends in 1733,[68] but it was only published five years later in the final edition of his book. This paper contains

---

where

$$t_j = 2pq\left[1 + \cos\left\{\frac{(2j-1)\pi}{b}\right\}\right]$$

and

$$c_j = \frac{\prod_{i\neq j}(1-t_i)}{\prod_{i\neq j}(t_j-t_i)}.$$

Note that the diagram corresponds to the case $b=10$ and that the lengths of the lines $QF$, $OE$, $MD$, $KC$ and $HB$ in the diagram are $\sin(\pi/10)$, $\sin(3\pi/10)$, $\sin(5\pi/10)$, $\sin(7\pi/10)$ and $\sin(9\pi/10)$, respectively. A reconstruction of De Moivre's solution exclusively based on tools available to him is contained in Schneider (1968, pages 288–292).

[67]In modern mathematical notation, the issue is to find the smallest number, $n$, of mutually independent Bernoulli trials $X_1,\ldots,X_n$ with common success probability $p$ for which given constants $c$ and $\alpha$, the event $\{|(X_1+\cdots+X_n)/n - p| \leq c\}$ occurs with probability greater than or equal to $1-\alpha$.

[68]This is De Moivre (1733), about which De Moivre (1756) writes in his preface: "I shall here translate a Paper of mine which was printed November 12, 1733, and was communicated to some Friends, but never yet made public, reserving to myself the right of enlarging my own thoughts as occasion shall require." A copy of the 1733 paper that originally belonged to James Stirling (1692–1770) is in the University of London Library. The inscription in De Moivre's handwriting reads simply: "for Mr. Stirling." References to early twentieth century discussions of De Moivre (1733) and the location of extant versions of it are in Daw and Pearson (1972). A modern reprint of the 1733 paper may be found in Archibald (1926), available on JSTOR.

larger, simpler approximations, which in turn lead to results that I am pleased to report below.

Let us suppose that there is an equal chance that an event may or may not happen, as for example, in the game of cross or pile,[69] and that the number of trials is arbitrary. As long as this number is greater than one hundred, the odds are then 28 to 13, or more than two to one,[70] that one of the cases will not occur more often than the other by more than half the square root of the latter number.[z] As the number of trials increases, the half of the square root decreases proportionally. This represents only the 120th part if it is 3,600, the 260th [sic] part if it is 14,400, the 2,000th if it is a million, and it vanishes

---

[69]"Cross and pile" refers to heads or tails on coins. Many early European coins had a cross on one side. Shown below, for example, is an English silver groat, a coin three pence in value, from the reign of King Edward III (1327–1377). The pile was the opposite or reverse side of the coin. It took its name from the under iron, called the pile, that was used in the minting apparatus to strike the coin. The die on the surface of the pile produced the reverse or pile side of the coin.

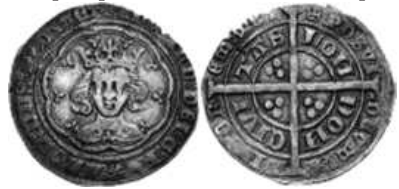

[70]Maty has given an abbreviated and garbled version of problems that appear in De Moivre (1733) to illustrate De Moivre's approximation to the terms in a binomial expansion. An English translation of De Moivre (1733) is in De Moivre (1738, pages 235–243 and 1756, pages 243–254). Assume $X$ is binomial with sample size $n$ and success probability $p=1/2$. De Moivre showed that for $n$ large relative to an integer $l$,

$$P\Big(X = \frac{n}{2} \pm l\Big) \cong \frac{2}{\sqrt{2\pi n}}\exp(-2l^2/n).$$

Crucial to this result is a form of the so-called "Stirling approximation" for $n!$ The latter was obtained by De Moivre independently of Stirling in 1730. In a series of corollaries, De Moivre used this approximation to obtain, for $c=1,2,3$,

$$P\Big(\Big|X - \frac{n}{2}\Big| \leq \frac{c\sqrt{n}}{2}\Big).$$

The resulting probabilities are given in terms of odds 28:13, 280:13 (the tenfold increase) and 369:1 for $c=1,2,3$, respectively. For details, see Schneider (1968, pages 296–299) and Schneider (2005). The odds are from De Moivre's own approximation to the probabilities given by 0.682688, 0.95428 and 0.99874, respectively; they may be compared to the probabilities resulting from the odds (0.6829, 0.9556 and 0.9973, resp.).

at infinity.[71] The size of the wager will increase tenfold if the range of the limits is doubled; it will be 369 to 1 if tripled, and considerably greater if multiplied tenfold. But were one to double, triple or multiply a hundredfold the range of these limits, it is possible to imagine a large enough number of trials that any connection with these limits will eventually disappear. The same calculations and arguments will apply in cases in which the probabilities of the events are in fixed relationships to one another.[72] Hence it follows that *in the long run, chance does not affect order*; *in other words, experience allows us to discover with certainty the results to which chance is subject*.[73]

According to our scholar,[74]

*We may imagine Chance and Design to be as it were in Competition with each other, for the production of some sorts of Events, and may calculate what Probability there is, that those Events should be rather owing to one than to the other.* [*To give a familiar Instance of this,*] *Let us suppose* [*that two Packs of Piquet-Cards being sent for, it should be perceived that there is, from Top to Bottom,*] *the same Disposition of the Cards in both Packs*; [*Let us likewise suppose that, some doubt arising about this Disposition of the Cards, it should be questioned whether it ought to be attributed to Chance, or to the Maker's Design*: *In this case, the Doctrine of Combination decides the Question, since it may be proved by its Rules, that*] *there are the Odds of above* 26,313,08[3] *Millions of Millions of Millions of Millions to One*,[75] *that the Cards were designedly set in the Order in which they were found.*

[*From this last Consideration we may learn, in many Cases, how to distinguish the Events which are the effect of Chance, from those which are produc'd by Design*:] *the very Doctrine that finds Chance where it really is, being able to prove by a gradual Increase of Probability, till it arrive at Demonstration, that where Uniformity, Order and Constancy reside, there also reside Choice and Design.*

In the dedicatory letter to Newton which prefaces the second edition of his book, Mr. De Moivre further wrote:[76]

*I should think my self very happy, if, having given my Readers a Method of calculating the Effects of Chance, as they are the result of Play, and thereby fix'd certain Rules, for estimating how far some sort of Events may rather be owing to Design than Chance, I could by this small Essay, excite in others a desire for prosecuting these Studies, and of learning from your Philosophy how to collect, by a just Calculation, the Evidences of exquisite Wisdom and Design, which appear in the* Phenomena *of Nature throughout the Universe.*

---

[71]The clear meaning of the two sentences beginning with, "As the number of trials increases..." and ending with, "...it vanishes at infinity" is that

$$\frac{1}{2\sqrt{n}} = \frac{1}{120}, \frac{1}{240}, \frac{1}{2000}$$

whenever $n$ is successively equal to 3,600, 14,400 and 1,000,000. This is a garbled attempt at explaining Remark I (e.g., De Moivre, 1756, pages 250–251) at the end of the section on De Moivre's approximation to the binomial. When $c = 1$ the probability in the previous footnote can be written as

$$P\left(\left|\frac{X}{n} - \frac{1}{2}\right| \leq \frac{1}{2\sqrt{n}}\right) = 0.682688,$$

or odds of about 2 to 1, as De Moivre says. De Moivre then notes that the fraction of the total number $n$ of cases that satisfies this probability is $1/(2\sqrt{n})$. Further, he calculates this fraction for the three cases that Maty gives, making the same typographical error of 260 instead of 240.

[72]In modern terms, Maty is saying that whatever the value of $c$, the probability of the event of interest converges to a fixed limit as $n \to \infty$, and the result continues to hold even when $p \neq 1/2$.

[73]This appears to be Maty's paraphrase of the last paragraph of Remark I following the normal approximation to the binomial distribution (De Moivre, 1756, page 251). The original reads "And thus in all Cases it will be found that altho' Chance produces Irregularities, still the Odds will be infinitely great, that in process of Time, those Irregularities will bear no proportion to the recurrency of that Order which naturally results from ORIGINAL DESIGN."

[74]This quotation is excerpted from the preface of De Moivre (1718, pages v–vi). The original passage is reproduced here, with brackets indicating the parts that Maty left out in his translation.

[75]The game of piquet had either 32 or 36 cards, depending on the version played. Here, De Moivre is considering the 32-card version, so that the probability of a perfect match between two such decks of cards would be 1 in 32! $\approx 26,313,083 \times 10^{28}$. It is interesting to note that in his attempt to make the magnitude of the probability easier to grasp, De Moivre ends up being off by a factor of $10^4$. In inadvertently dropping the last digit, Maty is off by an additional factor of 10.

[76]This quotation is taken verbatim from the first edition of *The Doctrine of Chances* (De Moivre, 1718); it is reproduced here in its original form. Maty's reference to "the second edition of his book" is presumably meant to say that he viewed *The Doctrine of Chances* as the second edition of *De Mensura Sortis*.

I felt it to be my duty to record these thoughts, which Mr. De Moivre communicated to me in person, adding that in his opinion, there was no more powerful argument against a system that would attribute the creation to a fortuitous collision of atoms, than that whose principles are set forth in his book.

I am uncertain whether to include among Mr. De Moivre's writings his revision of Mr. Coste's French translation of Newton's *Opticks*.[77] Recommendations made by the court had led the English philosopher to use the same hand as the one employed to translate [into French Locke's] *Essay on Human Understanding*. Now just as that hand had to be guided by Mr. Locke himself, it was fortunate to be assisted also in the present case by a mathematician trained by Newton himself; for otherwise, the essay would have been published with a plethora of errors, which Mr. De Moivre noticed immediately and corrected at Newton's bidding. The latter had absolute confidence in Mr. De Moivre for thirty years.[78] He took delight in his company and would arrange to meet him in a certain coffee-house[79] to which the French mathematician retired as soon as he had fin-

---

[77] Born in France and educated in Geneva, Pierre Coste (1668–1747) was another Huguenot refugee. He is known for his translation of several English works into French which helped introduce English thought to eighteenth century France. After Coste translated into French two works of the English philosopher John Locke (1632–1704), the latter invited him to England in 1697. There, he worked on the translation of Locke's *Essay Concerning Human Understanding* under the author's guidance. Following on this project, Coste subsequently worked as a tutor to the wealthy and the nobility. Coste translated the second edition of Newton's *Opticks* (Newton, 1718) into French (Newton, 1720). It was published in Holland. Another edition (Newton, 1722) was to be published in France. When it was submitted to the government censor for approval, the mathematician Pierre Varignon (1654–1722) was asked to look at the book (Newton, 1959–1977, Volume VII, pages xxxv–xxxvi, 200–201, 214–215). He not only approved of the publication but took charge of getting the work to print. From that point on, he was in contact with Newton about the publication. It is likely that Newton asked De Moivre to handle the corrections to the French edition and Coste was shunted to the side. Coste complained to Newton that his corrections were being ignored and that he had not been shown De Moivre's corrections as promised. Varignon did receive corrections from both De Moivre and Coste and commented that De Moivre's were more helpful. In the end, Coste acknowledged in the preface of Newton (1722) how De Moivre had improved the translation.

[78] Without giving any sources, Walker (1934) writes: "Tradition says that in his later years, Newton often replied to questions by saying 'Ask Mr. Demoivre, he knows all that better than I do.'"

[79] This was most likely Slaughter's Coffee-house in St. Martin's Lane, which was probably near where De Moivre lodged. A succinct description of the activities of a coffee-house is given in Lewis (1941, pages 32–33):

> "The coffee-house is where one may talk politics, read the ten London newspapers of the day, where one's letters may be addressed, where one makes appointments and where one may meet others of one's trade or profession."

Each coffee-house tended to have its own distinct clientele. According to Lillywhite (1963, page 530), Slaughter's was known as a meeting place for chess players as well as a place where Huguenots met. Prior to the establishment of the Royal Academy of Arts in 1768, it was also a meeting place for artists. Frequenting a coffee-house was probably ideal for De Moivre, whose lodgings may have consisted of only a couple of rooms. He definitely did not own or rent an entire house since his name does not appear in the Poor Law Rate Books for the City of Westminster. His lodgings were large enough, however, that he employed a servant by the name of Susanna Spella, whom he mentioned in his will (Public Record Office).

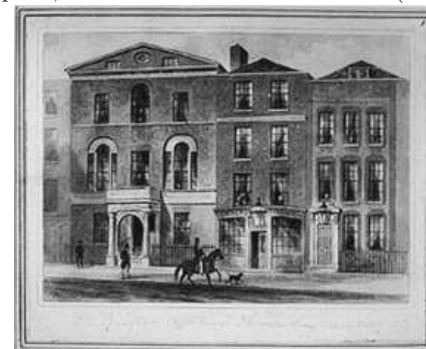

*Slaughter's coffee house*

There are at least three contemporary references that have De Moivre frequenting Slaughter's between 1712 and 1747. In a letter of October 12, 1712, De Moivre wrote to Johann Bernoulli that he should address his reply at Slaughter's Coffee-house (Wollenschläger, 1933, page 274). In a 1730 letter from Colin Maclaurin to James Stirling, Maclaurin mentions that he had written to De Moivre at Slaughter's Coffeehouse. The letter to De Moivre, which has also survived, was about Maclaurin's subscription for six copies of the *Miscellanea Analytica*. The letter, which was accompanied by the payment of the subscription, described who should receive the copies of the book (Maclaurin, 1982). In 1747, Jean-Bernard Le Blanc (1707–1781), the French abbot, author, historian and art critic, wrote a series of letters (Le Blanc, 1747) comparing France and England, their people and institutions. With regard to gambling, Le Blanc puts De Moivre at Slaughter's giving advice on gambling (Le Blanc, 1747, Volume II, page 309). Le Blanc also notes that De Moivre, although "the ished teaching.[80] Newton would take him back to his house, where they spent their evenings debating philosophical matters.[a]

The *Miscellanea Analytica*, which was published in 1730 and dedicated to Mr. Folkes,[81] the author's student and friend, is a compendium of his discoveries and methods. It contains derivations of the main theorems that Mr. De Moivre had stated without proof in his previous writings, particularly those concerning *recurrent series*. This book, intended as

---

greatest calculator of chances now in England," had never calculated the effects of gambling on morality (Le Blanc, [1747], Volume II, page 307).

There is also a more "modern" reference (Fiske, [1902]) to De Moivre playing chess at Slaughter's. Unfortunately, no sources were given in the publication for the information on De Moivre and some of the statements made by Fiske about De Moivre are inaccurate.

Lord Philip Stanhope, 2nd Earl Stanhope probably visited De Moivre at Slaughter's in 1744 (Centre for Kentish Studies). He recorded in his account book on July 24 that he paid a shilling at Slaughter's. If they did meet at Slaughter's that day, it was to discuss mathematics. Earlier, on July 5 and 12, De Moivre had written to Stanhope (Centre for Kentish Studies). The earlier letter begins: "Since I had the honour a seeing your Lordship..." The subjects of the letters, as well as an undated third letter, were a topic from the *Miscellanea Analytica* and a result due to Euler.

Slaughter's Coffee-house was not the only one that De Moivre patronized. Edward Montagu (1678–1761), a former De Moivre student and at the time Member of Parliament for Huntingdon, wrote to his wife in 1751 (Climenson, [1906]):

"I desire when wheatears are plenty and you send any to your friends in London, you would send some to Monsieur de Moivre at Pons Coffee House in Cecil Court in St. Martin's Lane, for I think he longs to taste them."

Pons coffee-house was frequented by the more prominent Huguenots or, as quoted by Lillywhite ([1963], page 450) from an original source, some "foreigners of distinction."

[80]For much of his career, De Moivre tutored the sons of the wealthy and titled in order to make a living. One of his earliest aristocratic clients was William Cavendish (1641–1707), 1st Duke of Devonshire. De Moivre probably paid more than his respects to the Earl of Devonshire (later Duke) as noted in Maty's anecdote of De Moivre seeing the *Principia Mathematica* for the first time. Maty's list of De Moivre's students includes a Cavendish, probably Lord James Cavendish, a younger son of the Duke. The eldest son was probably also a student; the 2nd Duke subscribed to the *Miscellanea Analytica*. The role of tutor probably continued into another generation; a younger son of the 2nd Duke is also on the subscription list. Another aristocratic client was Ralph Montagu, 1st Duke of Montagu; De Moivre gave lessons of mathematics to the Duke's son, John Montagu, later 2nd Duke of Montagu (Murdoch, [1992]).

Within a decade of his arrival in London, De Moivre had become well established as a mathematics teacher. Early in 1695, there was an attempt to establish via a lottery two Royal Academies that would provide instruction in languages, mathematics, music, writing, singing, dancing and fencing. An advertisement in the February 22, 1694/5 issue of the journal *A Collection for Improvement of Husbandry and Trade* shows Abraham De Moivre and Richard Sault (d. 1702) as the two mathematics teachers (Anonymous, [1695]). De Moivre continued to teach mathematics throughout his career, as evidenced by a poem of Deslandes ([1713]) in which De Moivre is referred to as an "eminent teacher of mathematics."

On his arrival in England De Moivre apparently tried his hand, unsuccessfully, at lecturing in coffee-houses. The Penny Cyclopaedia states:

"He appears at the earliest period to which any account of him reaches to have devoted himself to teaching mathematics, as the surest means of obtaining a subsistence. He also, though he was not the first who adopted that plan, read lectures on natural philosophy: but it does not appear that his attempts in this way were very successful, he neither being fluent on the use of the English language, nor a good experimental manipulator." (Society for the Diffusion of Useful Knowledge, [1837], page 380).

By the time of the publication of *The Doctrine of Chances* in 1718 his written English, at least, had become very good.

[81]In the subscription list to the *Miscellanea Analytica*, Martin Folkes (1690–1754) is listed as having ordered seven copies. At his death, Folkes still possessed three copies of the book in his library; they were in various bindings and types of paper (Baker, [1756]). Also on the subscription list are Martin's brother, William Folkes (ca. 1700–1773) and uncle, Thomas Folkes (d. 1731). Martin and William's father, also Martin Folkes, died in 1706. It is probable that their uncle Thomas arranged for them both to be taught mathematics by De Moivre.

The strength of the friendship, as well as the professional connection, between Martin Folkes and Abraham De Moivre might be guessed from what little historical information survives. Folkes had copies in his library of all editions of *The Doctrine of Chances* and *Annuities upon Lives* with multiple copies of some of the editions. In addition, he had a mathematical manuscript by De Moivre that commented on Newton's *Quadrature* (Baker, [1756]). There are two recorded visits between De Moivre and Folkes. They dined together in 1747 on the occasion of De Moivre's eightieth birthday; also in attendance was Edward Montagu, another of De Moivre's former pupils (Stirling and Tweedie, [1922]). Sometime, perhaps late in his life, De Moivre visited Folkes at his house. There is a letter (Royal Society, Folkes Collection) in French from De Moivre to Folkes asking if he could make a short visit to Folkes that day. The hand is uneven and so the note was possibly written in old age.

it is for only the very best mathematicians,[82] is uncommon inasmuch as the propositions contained therein are presented separately from their proofs in order to allow the mind to grasp the logical connections more easily, while spurring it to independent discovery of the proofs.

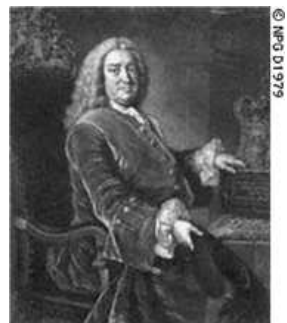

*Martin Folkes*
*1690–1754*

Mr. Naudé,[83] the famous mathematician from Berlin, was provided by Mr. De Moivre with a copy of this book, along with a letter containing the solutions to several algebraic problems for him to present it at the Berlin Academy of Sciences.[84] At the Assembly of August 23, 1735, he tabled a proposal that a man of such great distinction should be appointed a member.[b] The proposal was put to the vote and Mr. De Moivre's election was ratified by a kind of acclamation.

The publisher of Jacques Bernoulli's book[85] invited Mr. De Moivre to follow the example of this famous writer by applying the science of probabilities to daily life. Our scholar politely declined to undertake this new task. However, the invitation seems

---

[82] The *Miscellanea Analytica* contains the only extant subscription list for any of De Moivre's books. There was a subscription to *The Doctrine of Chances* (De Moivre, 1718), but the list of subscribers was not printed. All editions to *Annuities upon Lives* were probably not sold by subscription. An advertisement in Wilford (1723–1729, Volume II) states that the first edition (De Moivre, 1725) could be obtained from two different booksellers, Francis Fayram at the Royal Exchange and Benjamin Motte at Temple Bar, at a cost of three shillings.

In an advertisement in Wilford (1723–1729, Volume III), there is a description of how De Moivre put together the subscription list. He contacted several people himself, probably by letter, and took payment for their subscriptions. He then advertised that he was printing a few more copies than there were subscriptions so that anyone wanting a copy should contact a bookseller in St. Martin's Lane near where he lived. The cost for the subscription was one guinea, or 21 shillings.

There are some very astute mathematicians on the subscription list, but they are in the minority and so there must be other explanations for buying the book. The mathematicians include William Jones (1675–1749), Samuel Klingenstierna (1698–1765), Colin Maclaurin (1698–1746) who ordered six copies for himself and his friends, John Machin (1680–1751) and Pierre de Maupertuis (1698–1759). Gabriel Cramer (1704–1752) also ordered a copy through the bookseller William Innys. There were others who were amateur mathematicians. The mathematicians are, however, a small minority of approximately 160 subscribers in total, excluding some college libraries from Cambridge. The complete list of subscribers includes members of the aristocracy [including the 2nd Duke of Montagu, a known patron of Huguenots (Murdoch, 1992), who bought ten copies] and their relations, members of Parliament, fellows of the Royal Society and some Huguenot friends. The aristocracy and the parliamentarians on the list were mostly Whigs by political persuasion. Several subscribers had a fairly close connection to Isaac Newton including John Conduitt, the husband of Newton's niece, who bought 15 copies. Some subscribers were probably De Moivre's former students. Bellhouse, Renouf, Raut and Bauer (2007) has analyzed the subscription list and has suggested that one of the main themes behind the act of subscribing in this case is the provision of patronage for the new Euclid of probability, the man who had systematized chance. A poem in praise of De Moivre (Deslandes, 1713) begins by calling him the new Euclid.

[83] Born in the French city of Metz, Philippe Naudé (1684–1745) became Professor of Mathematics at the Royal College of Joachim in Berlin (Formey, 1748, pages 465–468). The family fled to Berlin after the revocation of the Edict of Nantes in 1685. De Moivre returned the favor that was given to him. Naudé was elected Fellow of the Royal Society in 1737; De Moivre was one of his sponsors with Martin Folkes, De Moivre's friend, the first sponsor (Royal Society EC/1737/17).

[84] Founded in 1700 by Frederick III, Elector of Brandenburg, with Leibniz as its first president, the Academy was known originally as the "Berlin-Brandenburgische Sozietät der Wissenschaften" (Berlin-Brandenburg Society of Scientists). In 1743, the academy was reorganized under Leonhard Euler with the new name "Académie royale des sciences et belles lettres" (Knobloch, 1998). Its present name is Berlin-Brandenburgische Akademie der Wissenschaften.

[85] The publisher of Bernoulli's *Ars Conjectandi* is given in Latin as Thurnisiorum Fratrum. This refers to the brothers Emmanuel and Johann Rudolph Thurneysen. Note, however, that Maty is wrong in his statement that it is these publishers who encouraged De Moivre to write on these subjects. The invitation came from Nicolaus Bernoulli, who edited his uncle's *Ars Conjectandi*. In the preface to the book, he asked both De Moivre and Montmort to consider economic and political applications of probability, subjects that his uncle Jacob Bernoulli (1654–1705) had intended to pursue.

to have induced him, in 1721, to initiate new research on probabilistic issues connected with human life. England is probably the country where such matters as the value of life annuities, substitution contracts, purchases of expectations[86] and so on, are most common. Prior to Mr. De Moivre's work, the English blindly followed the same incorrect and customary recipes. Thus, our Islanders enthusiastically welcomed the simple, general and precise rules that Mr. De Moivre put forward in his *Annuities upon Lives*, published initially in 1724[87] and again in 1743. As the theory on which his techniques are based is strictly his own, I cannot gloss over any details lest I distort them by trying too hard to be brief.

As early as 1692, Dr. Halley had drawn up a mortality table based on the Breslau registers.[c] He had even developed some rules for calculating annuities for one or more lives. However, the calculations for each single life involved as many arithmetic operations as there were years between a person's current age and the point at which this person turned a hundred. When it came to calculating the sums and differences for several lives, there was a phenomenal rise in the number of combinations; and even the inventor agreed that, despite the convenience of logarithms, it was preferable to find a shorter method than his own. It would not be easy to find what Halley had sought in vain. Nonetheless, Mr. De Moivre applied himself to the task and his results exceeded his expectations. He began by observing—it is surprising that Halley had not seen this himself—that there were intervals of several years during which the length of human life decreases uniformly. Of 646 adults of 12 years of age, namely the survivors of childhood mortality out of an initial group of one thousand, six die every year, twelve every two years and so on, up to age 25. Each of the subsequent four years, seven more die. From ages 29 to 34, the annual proportion is eight; it is then nine up to age 42, ten up to age 49, and eleven up to age 54.[d] The proportion drops back to ten up to age 70, rises to eleven again up to age 74, and returns to ten up to age 78. The death rate then follows an arithmetic progression of nine, eight, seven and six for the four subsequent years, and of the twenty people still living at age 86, one at most will live to one hundred. Mr. De Moivre was not content with his discovery of these intervals, which alone shorten the time of calculation considerably; he further observed that their inequalities balance each other. He thus concluded that they can be regarded as parts of an arithmetic progression that could be computed with more abundant, accurate data.[88] The first term of this progression may be set at age 12, and the last one at age 86. Of 74 adolescents of the former age, one must die every year, and the interval between their individual ages and the time they die is their complement of life. Each age corresponds to a series, which expresses the probability of life expectancy; when multiplied by the amount of the individual's life annuity for that number of years, it represents the value of the annuity. Mr. De Moivre had no difficulty calculating this value and consequently produced a very simple formula that could be applied whatever a person's age. It requires just four easy operations, and anyone with a basic knowledge of arithmetic can per-

---

[86]The term "life annuities" is still in common use today, but the others are not. A substitution contract probably refers to leases based on the lives of the lessees: on the death of one of these lessees, another person could be substituted into the lease through a monetary payment whose value needed to be determined. As for the purchases of expectations, they likely refer to reversionary annuities, as exemplified in Problem XXVII of De Moivre (1725).

[87]The date of publication for the first edition is 1725 (De Moivre, 1725).

[88]Pearson (1978, pages 146–154) examined De Moivre's piecewise linear solution in detail. He looked at Halley's data and concluded that "De Moivre's hypothesis deviates considerably from the truth." He also noted that this may not be important if the hypothesis provides a reasonable approximation to the price of an annuity. For a life age 50 and using 5% interest, Pearson found that the price of the annuity using De Moivre's method was slightly greater than 4% over the price without the approximation. The approximation then would be in favor of the annuity vendor. These calculations were done, either by hand or hand calculator, by an actuary that Pearson knew. The fact that the actuary did the calculations for one special case only points to the enormity of the burden of calculation for annuity valuations done by hand in the eighteenth century.

|      | Age at issue |      |      |     |     |     |     |     |     |     |     |
|------|------|------|------|-----|-----|-----|-----|-----|-----|-----|-----|
| Rate | 20   | 25   | 30   | 35  | 40  | 45  | 50  | 55  | 60  | 65  | 70  |
| 3%   | −3.8 | −2.0 | −0.3 | 1.2 | 2.5 | 3.8 | 4.6 | 4.3 | 5.3 | 6.8 | 8.9 |
| 5%   | −3.7 | −2.2 | −0.7 | 0.7 | 1.9 | 3.1 | 4.0 | 3.7 | 4.8 | 6.3 | 8.8 |
| 7%   | −3.6 | −2.3 | −0.9 | 0.3 | 1.4 | 2.6 | 3.5 | 3.2 | 4.3 | 5.9 | 8.7 |

The table above shows, for various rates of interest and ages at issue, the percentage increase over the true price of the annuity when De Moivre's approximation is used instead of the complete set of calculations using Halley's life table. At younger ages, the approximation is in the annuitant's favor.

form this calculation with the help of appropriate tables. The same rules apply to joint lives, survivors and mortgages and so on. Indeed, our mathematician's rules are so simple *that by the Help of them, more can be performed in a Quarter of an Hour, than by any Method before extant, in a Quarter of a Year.*[e]

However, annuities computed in this manner are subject to the following condition: payment is due every year and if the holder dies, the payment for the year of his death is forfeited by his inheritors. When this condition is changed so that the payments cease at the very moment of death, a different problem arises for which our mathematician proposed a solution in a memoir that he communicated to the Royal Society in 1744. He also demonstrated therein how the various intervals of a person's life should be linked and how their probabilities should be computed, on the basis of the data alone. As one of his students has shown,[f] the accumulation of data tends to confirm Mr. De Moivre's general formula. Furthermore, the simplicity of nature is grounds for believing that yet again, he has uncovered a rule that ultimately transcends chance, though subject it may be to anomalies in a few cases.[89]

Mr. De Moivre's life was as uneventful as it was rich in discoveries and writings. To a certain extent, it could be compared to a sequence in which each term encompasses and is greater than those which precede it. It is regrettable that such a sequence should have a final term and that a man who enriches society daily through his students, and who enhances science through the discoveries he makes, cannot be freed from the limits of the human condition. Nonetheless, there is a difference between Mr. De Moivre in the latter stages of his life and the common run of men: although the faculties of his soul became less resilient, they lost none of their vigour. He suffered partial loss of sight and hearing[90]; his body required more rest and his mind, greater respite. Although he came to need twenty hours sleep, he spent the remaining three or four hours taking his only meal of the day and talking with his friends. For the latter, he remained the same: always well-informed on all matters, capable of recalling the tiniest events of his life, and still able to dictate answers to letters and replies to inquiries related to algebra.

It was during this last period of a life reduced to its smallest terms—if I may be allowed to refer to a mathematician in this way—that he learned that he had been admitted to the Royal Academy of Sciences in Paris.[91] He was overjoyed and declared on several occasions that he regarded this election as the crowning moment of his career. In a letter to Mr. De Mairan,[92] which he found the energy to dictate and sign, he expressed his commitment and gratitude with enthusiasm. However, he overestimated the time that he probably had left to live and underestimated the difficulty of recovering the manuscripts that he had lent when he promised to

---

[89] Maty has left out completely De Moivre's dispute with Thomas Simpson (1710–1761). Briefly, the fight was about Simpson's incursion into De Moivre's domain of expertise with books that were for the most part simplifications and popularizations of De Moivre's work on probability and annuities. Schneider (1968, page 216), Stigler (1986, pages 88–90) and Pearson (1978, pages 170–182) describe the dispute in detail. Simpson supported himself in part by writing inexpensive textbooks, the first of which was a book on integral and differential calculus (Simpson, 1737). Initially, relations between De Moivre and Simpson were cordial. This changed, however, after the publication of Simpson's next two books, one on probability (Simpson, 1740) and one on annuities (Simpson, 1742). In the preface to the second edition of his book on annuities, De Moivre (1743, page xii) complained:

> "After the pains I have taken to perfect this Second Edition, it may happen, that a certain Person, whom I need not name, out of *Compassion to the Public*, will publish a Second Edition of his book on the same Subject, which he will afford at a *very moderate Price*, not regarding whether he mutilates my Propositions, obscures what is clear, makes a Shew of any Rules, and works by mine; in short, confounds, in his usual way, everything with a crowd of useless Symbols; if this be the Case, I must forgive the indigent Author, and his disappointed Bookseller."

Simpson quickly replied that De Moivre's behavior was ungentlemanly. De Moivre was tempted to make one more riposte but was dissuaded by his friends.

[90] The advertisement to De Moivre (1756, page xi) rather refers to the "failure of Eye-sight" and in his eulogy of De Moivre, Fouchy (1754) writes that "he found himself successively deprived of sight and hearing." [authors' translation]

[91] There was a fixed number of foreign members in the Académie royale and new members were admitted only to replace those whose memberships terminated by death. When Prussian philosopher and mathematician Christian Wolf (1679–1754) died, the Académie, at their meeting of August 14, 1754, put forward two names for consideration to the king: Abraham De Moivre and Swiss biologist Albrecht von Haller (1708–1777). The Académie was informed three days later that the king had chosen De Moivre (Bibliothèque nationale de France).

[92] This is again Jean-Jacques Mairan, who had now become "secrétaire perpétuel de l'Académie."

repay the honor bestowed upon him through some scholarly tribute.

He was to enjoy this recognition for a few months only. His health grew steadily worse and he needed to sleep longer and often. After being confined to bed for seven or eight days, he died in his sleep on November 27, 1754.[93]

It behoves those people qualified to read Mr. De Moivre's writings to assign him his place in history. The rest may judge him by the friends he had and the students he trained. Newton, Bernoulli, Halley, Varignon, Sterling, Saunderson, Folkes and many others could be listed in the first group; Macclesfield, Cavendish, Stanhope, Scot[t], Daval[l] and Dodson,[94] belong to the second.[95]

Had it not been for his need to give lessons, he would no doubt have risen to even greater heights. Efforts were made on his behalf to free him from his state of dependence by obtaining a professorship for him at the University of Cambridge.[96] However, he was a foreigner, and frankly, he lacked the kind of savvy needed to win the favour of those who could have ensured that his origins be forgotten and his talent recompensed.[97]

His knowledge extended beyond the purview of mathematics. His love of humanities and belles-lettres remained constant. He was keenly aware of the beauty of the classics and was often consulted on obscure and controversial passages from these works. His favourite French authors were Rabelais and Molière,[98] and he could recite them by heart. He once told one

---

[93] De Moivre was buried four days later from St. Martin-in-the-Fields church on December 1, 1754 (Westminster Council Archives).

[94] Bellhouse (2007a) argues that the students on the list are: George Parker (1697–1744), 2nd Earl of Macclesfield; probably Lord James Cavendish (1673–1751), third son of the 1st Duke of Devonshire, or possibly Lord Charles Cavendish (1693–1783), second son of the 2nd Duke of Devonshire; Philip Stanhope, 2nd Earl of Stanhope (1714–1786); George Lewis Scott (1708–1780); Peter Davall; and James Dodson (1709–1757), respectively. Augustus De Morgan (1806–1871) had a different interpretation for one of the names on the list (De Morgan, 1857). He assumed that "Stanhope" meant Philip Dormer Stanhope, 4th Earl of Chesterfield. Bellhouse (2007a) has argued against the Chesterfield interpretation based on Earl Stanhope's mathematical background and the Earl of Chesterfield's lack of interest in mathematics. De Morgan is not the only eminent mathematician to have mixed up Philip Stanhope and Philip Dormer Stanhope. Pearson (1978) assumed incorrectly that it was the latter Stanhope who nominated Bayes for fellowship in the Royal Society.

[95] Barnard (1958) has speculated that Thomas Bayes (1701?–1761) was another of De Moivre's students, writing that "Bayes may have learned mathematics from one of the founders of the theory of probability." This is unlikely. Bayes studied at the University of Edinburgh, probably learning his mathematics from the professor of mathematics at Edinburgh at the time, James Gregory. Bellhouse (2007a) has suggested that it was Philip Stanhope who initially met Bayes and got him interested in working on problems in probability.

[96] The position came open in 1739 on the death of the Lucasian Professor of Mathematics, Nicholas Saunderson (1682–1739). There were two candidates for the position, De Moivre and John Colson (1680–1760). Ball (1889, page 101) has described the election succinctly as follows:

"When a candidate for the Lucasian chair in 1739, he [Colson] was opposed by Abraham de Moivre, who was admitted a member of Trinity College and created M.A. to qualify him for office. Smith [Robert Smith, the master of Trinity College] really decided the election, and as de Moivre was very old and almost in his dotage he pressed the claims of Colson. The appointment [of Colson] was admitted to be a mistake ..."

The Cambridge University registers (Venn and Venn 1922–1954) show De Moivre obtaining an M.A. in 1739.

[97] In De Moivre's eulogy, Fouchy (1754) turns the sentence into a double-entendre by writing "However, he was a foreigner, and frankly, he lacked the kind of savvy needed to win the favor of those who could have ensured that this quality be forgotten." [authors' translation] Here, the word "quality" could be taken neutrally as in "condition" but also positively as an "advantage," which might be interpreted as a snub at the English scholarly elite.

[98] According to another source (Motteux, 1740, page 114), De Moivre enjoyed reading the French authors Corneille, Molière, La Fontaine and Rabelais. Motteux (1740) is a posthumous edition with several footnotes added by César de Missy (1703–1775), who was French Chaplain to King George III. In one footnote, de Missy remarks that there was some question over whether Book V of Rabelais's *Gargantua* was actually written by him. De Moivre, among others, not only attributed the book to Rabelais but deemed it to be the best part of the work. Le Blanc (1747, Volume I, page 155) describes De Moivre as "not less a lover of the elegant arts than of geometry." Pierre Coste (Montaigne, 1754, Volume IV, page 133) also notes De Moivre's familiarity with Montaigne's *Essais*.

De Moivre also read contemporary commentaries on French literature and the arts. For example, he expressed interest in receiving the 1740 edition of Jean-Baptiste Dubos's reflections on poetry and painting. Dubos was the secretary to the Académie française (Le Blanc, 1747, Volume I, page 155).

There are also possible connections to English literary society. The celebrated English poet Alexander Pope (1688–1744) included a reference to De Moivre in his epic poem *An Essay on Man* (Pope, 1734). The relevant lines in the poem are:

"Who made the Spider Parallels design,
Sure as De-Moivre, without rule or line?"

of his friends that he would rather have been Molière than Newton. He recited scenes from *Le Misanthrope* with all the flare and wit that he recalled seeing them presented with on the day he saw the play performed in Paris 70 years earlier by Molière's own company. It is true that misanthropy was nothing new to him.[99] He was a stern judge of men and at times, a glance was all that was required for him to form a judgment. He was unable to conceal sufficiently his impatience with stupidity and his hatred of hypocrisy and lies.

His discourse was far-reaching and instructive.[100] He never tried to flaunt his knowledge, and he showed himself to be a mathematician simply through the soundness of his mind. He was lucid and methodical in his conversation, his teaching and his writing. He only spoke after careful thought. Strength and depth rather than charm and liveliness were the hallmarks of his conversation and writing. His English and Latin essays were models of concision and accuracy. He devoted equal time and energy to polishing his style as he did to refining his calculations, and it is a testament to his perseverance that one is hard pressed to find errors in any of his work.

He understood the cost and importance of time only too well to waste it.[101] Nor did he allow matters of idle curiosity to distract him from his purpose. On one occasion, he declined to answer a friend's question as it entailed a huge set of calculations and did not in his opinion deserve his time and attention.

---

The quotation is from the third epistle about the growth of society. It is impossible to say whether Pope knew De Moivre or just knew of him.

De Moivre was also interested in music (possibly through his brother Daniel), or at least the mathematical aspects of it. When the composer and music theorist Johann Pepusch (1667–1752) tried to work out the mathematical theory behind ancient Greek music, he consulted De Moivre and his student George Lewis Scott. De Moivre "used to call him [Pepusch] a stupid German dog, who could neither count four, nor understand any one that did" (Burney, 1789, Volume IV, page 638). The comment may have been made only in jest; De Moivre and Scott were two of Pepusch's sponsors for fellowship in the Royal Society (EC/1745/09). Pepusch eventually published his insights into ancient Greek music in the form of a letter to De Moivre (Pepusch, 1746). De Moivre made other strong comments about his contemporaries, again perhaps in jest. In a letter (Columbia University) to Edward Montagu, De Moivre referred to Henry Stewart Stevens (d. 1760) as a fool since the latter could not solve or even begin a challenge problem in probability. The letter probably dates from the mid-1720s; in 1740 De Moivre was the first proposer for Stevens's fellowship in the Royal Society.

De Moivre's brother, Daniel, was an accomplished flautist. He composed, taught and performed on the instrument. In 1695, there was an attempt to found some Royal Academies to provide instruction in the arts and sciences. Abraham De Moivre was one of the proposed instructors in mathematics and Daniel an instructor of the flute or recorder (Tilmouth, 1957). Between 1701 and 1715, Daniel composed and published three collections of music for the recorder (Stratford, 1987). He also performed at Stationers Hall, one of the leading musical venues in London, as well as at taverns and coffeehouses (Lasocki, 1989).

[99]"Le Mysanthrope" is one of Molière's most famous plays; it was first performed on June 4, 1666. Maty's wording suggests a parallel between De Moivre and the hero of the play. Having lost all patience with the flattery and hypocrisy of fashionable society, the latter has vowed to speak and act only with complete sincerity. Paradoxically, he falls in love with the epitome of all that he despises, a cruel coquette. Disgusted by his loss in a lawsuit in which justice was on his side, he resolves to abandon society once and for all, and asks his true love to accompany him. Unfortunately, she is more in love with her frivolous lifestyle than with him. In the end, the hero departs alone.

[100]Charles-Étienne Jordan (1700–1745) visited De Moivre in 1733 (Jordan, 1735, pages 147 and 174). He described De Moivre as a man of wit and of pleasant company. Jordan, the son of Huguenot refugees, was born in Berlin and worked for Frederick the Great of Prussia. When Jordan's wife died in 1732, he fell into a depression and was counselled by his family to travel. He decided to go to France, Holland and England to meet some of the leading literary and scientific figures which included Voltaire in France, mathematicians Willem s'Gravesande and Pieter van Mussenbroeck in the Netherlands, as well as Alexander Pope and Abraham De Moivre in England (Frederick II, 1789, pages 5–7).

[101]There is one example of where De Moivre may have wasted his time. It concerns a proposed method to measure longitude at sea. The measurement of longitude had been such an important practical problem that in 1713, the British Parliament offered a prize of £20,000 for its solution. The responsibility for awarding of the prize fell to the Commissioners of Longitude. The only woman to try for the prize was Jane Squire. In 1731, she proposed a method to divide the sky into more than a million numbered spaces, which she called "cloves." Based on the clove directly above the navigator at sea, and using an astral watch that was set to the movement of the stars, the navigator could calculate the longitude from Squire's prime meridian which ran through the manger at Bethlehem. In 1742, Squire published her correspondence (Squire, 1742) with the Commissioners and other scientists; they were all sceptical of her method. De Moivre was one of her correspondents. From a friend, he had learned that her proposal was based on the exact course of the ship and the distance traveled by the ship. He pointed out that in practice, these measurements were very imperfect. Squire replied that De Moivre had been misinformed by his friend and that her method was based on using the fixed stars.

When a beloved nephew[102] of his passed away some time later, however, he did return to the problem and solved it since it distracted him from his grief.

Those who claim to have surmised his beliefs deem that his faith did not extend beyond Naturalism, but they maintain that his scepticism was in no way absolute, that he regarded religious revelation as an enigma, and that he could not suffer people who leveled unfounded charges or treated such questions with derision. One day, he said to a man who had blamed mathematicians for their lack of faith: *I'm going to show you that I'm a Christian by forgiving the inane remark you have just made*!

Mr. De Moivre never took a wife. Mathematics did not make him rich and he lived a mediocre life,[103] bequeathing his few possessions to his next-of-kin.[104]

---

[102]His nephew Daniel De Moivre died in July of 1734 and his brother, also Daniel De Moivre, less than a year earlier in September of 1733 (Public Record Office).

[103]Literally, "son état a été de la médiocrité." In eighteenth century French and English, "mediocre" meant "average to below average" whereas it is usually taken to mean "poor" in modern days. This is perhaps the reason why De Moivre is generally described as having "died in poverty" in contemporary sources. In his eulogy of De Moivre, Fouchy (1754) combines this sentence with the previous one to state that "The mediocrity of Mr. Moivre's fortune made it impossible for him to ever consider getting married." [authors' translation]

[104]Throughout his life, De Moivre failed to obtain any kind of patronage appointment that would allow him to pursue his research interests and live comfortably. Le Blanc (1747, Volume I, pages 168–169) comments on the situation by comparing De Moivre to the famous castrato singer Farinelli. After noting that Farinelli made large amounts of money on the stage, Le Blanc comments of De Moivre:

> "...it is surprising that a gentleman, who has rendered himself so valuable to science which they [the English] honour most, that Mr. De Moivre one of the greatest mathematicians in Europe, who has lived fifty years in England, has not the least reward made to him; he, I say, who, had he remained in France, would enjoy an annual pension of a thousand crowns at least in the academy of sciences."

There is a reference to De Moivre's "poverty" in the 1710s in correspondence between Leibniz and Bernoulli (Leibniz, 1962). The reference to poverty may have been made in comparison to patronage or university appointments, as enjoyed by Leibniz and Bernoulli, respectively.

However, De Moivre was not particularly poor when compared to the general population. When he died in 1754, he left £1600 in South Sea Annuities to his grandnieces Sarah and Marianne De Moivre, grandchildren of his brother Daniel.

The legacy was akin to a government annuity, or more specifically a perpetuity; the South Sea Company had taken over part of England's national debt and the money was raised through sale of shares and annuities. The speculation on shares went rampant and ended in the South Sea Bubble. Other evidence of De Moivre's lack of poverty is the free distribution of some copies of his books. As noted already at various points here, the Earl of Sunderland and Montmort received *De Mensura Sortis* and Johann Bernoulli received the *Animadversiones* concerning Cheyne. De Moivre also supplied continental mathematicians with English mathematical books by other authors, sometimes without expecting reimbursement. For example, in a letter from Pierre Varignon to Isaac Newton in 1722, Varignon writes (Newton, 1959–1977, Volume VII, page 209):

> "I beg you to pay Mr. De Moivre, on my behalf, the price of the posthumous book of Mr. Cotes (Cotes, 1722), which he recently sent me: I shall deduct the sum from the expenditure made and to be made by me on your account, as soon as I learn how much it is in our money."

According to King (1804, pages 48–49), persons in the sciences and liberal arts were making about £60 a year. Where did De Moivre's money come from? Teaching would not have brought in large quantities of money. Sales of his books may not have amounted to much either. One of his more popular books, the first edition of *Annuities upon Lives* sold for three shillings a copy. A normal book run of 500 copies would have amounted to £75 gross and much less net. It is likely that he received small patronage amounts from many of his aristocratic friends and clients. He also did some consulting, on issues related to his work both in annuities and probability. Schneider (2001) has made reference to an item, in a Berlin archive, where one can find answers by De Moivre to a client about financial mathematics. Fitz-Adam (1755–1757, Volume I, page 131), which is a collective pseudonym for Edward Moore, Lord Chesterfield and several others, has made reference to calculations that De Moivre did for someone regarding the ratio of married women to married men based on the Bills of Mortality. His advice with respect to gambling is found in at least two sources. An anonymous writer (Anonymous, 1731, page 8) referred to gamblers versed in mathematics and the calculation of chances as "de Moivre men." More telling of De Moivre's actual work in this area, Le Blanc (1747, Volume II, page 307) recounts:

> "I must add that the great gamesters of this country, who are not usually great geometricians, have a custom of consulting those who are reputed able calculators upon the games of hazard. M. de Moivre gives opinions of this sort every day at Slaughter's coffee-house, as some physicians give

His manuscripts are in the hands of a few friends,[105] equally well known for their erudition as they are for their determination to preserve his heritage. They alone are responsible for publishing whatever may still be of value in his work, and their own merit is so great that they could not possibly deprive others of materials capable of enhancing their life and times.

## ACKNOWLEDGMENTS

Funding in partial support of this work was provided by the Natural Sciences and Engineering Research Council of Canada, and by the Fonds québécois de la recherche sur la nature et les technologies. The authors are grateful to: Professor Alan Manning (Université Laval) for assistance and advice with the English translation of Maty's biography; Professor Ivo Schneider (Universität der Bundeswehr München) for many helpful comments on an earlier draft of the paper; Professor Duncan J. Murdoch (University of Western Ontario) for carrying out some library searches at the University of Oxford; and Professor Stephen M. Stigler (University of Chicago) for providing a copy of the allegorical picture from De Moivre's *The Doctrine of Chances*.

## MATY'S FOOTNOTES TO HIS BIOGRAPHY OF DE MOIVRE

[a] The memoirs that Mr. De Moivre dictated to me a few weeks before his death end here.

[b] *Commerc. Epistolic.* vol I, page 464. See also the *Leipzig Proceedings*, 1699, page 585. Mr. Facio disavowed this relationship as being false and without merit. *Comm. Epist.* vol. II, page 29. [The reference to *Commerc. Epistolic.* and the other abbreviation is to Leibniz's correspondence published in two volumes in 1745. The full title is *Commercii Epistolici Leibnitiani*.]

[c] Perhaps his nocturnal habits explain an amazing occurrence that the sceptical mathematician related to a few friends. One day, as he was working at a very early hour in his study, his mind was suddenly filled with light, causing him to make significant discoveries concerning the probabilities he was investigating. He said that this light, which remained with him

    their advice upon diseases at several other coffee-houses about London."

[105] In his will, De Moivre left his manuscripts to one of his former students, George Lewis Scott, who was also one of the executors of the will.

for several days, could well be construed by some people as a kind of inspiration.

[d] This is taken from the registers of the Royal Society. Mr. Birch was kind enough to check this for me.

[e] See the *Miscellanea Analytica*, page 88.

[f] See *Commerc. Epist.*, vol. I. page 462 and vol. II, page 11 and the *Leipzig Proceedings* of May 1700 with Mr. Moivre's memoir in the *Philosophical Transactions* 1702, no. 278, page 1126.

[g] See the Memoirs published that year, page 529, as well as the *Leipzig Proceedings* of March 1713. As of 1708, Mr. Keil had attributed this problem to the discoverer. As is apparent in the *Philosophical Transactions*, no. 317, he credited him with this honour in his writings published in the *Journal littéraire*, vol. VIII, page 420, and vol. X, page 181. The replies by Mr. Crufius on this matter are contained in the *Leipzig Proceedings* of October 1718. [The *Leipzig Proceedings* are known as *Acta Eruditorum*.]

[h] *Miscellanea Analytica*, page 17.

[i] It can be found in his *Miscellanea*, ibid.

[k] See his *Œuvres*, vol. IV. pages 67–68.

[l] *Commerc. Epistol.*, vol. II, page 187 & page 222.

[m] See the letter to Abbot Conti in Mr. Des Maiseaux's *Recueil*, vol. II, page 10. [The author is Pierre Des Maizeaux—note the variant spelling.]

[n] L'Evêque's oldest son. He was known personally to, and much esteemed by, Mr. Leibnitz and Mr. Bernoulli. He is often mentioned in their correspondence.

[o] Minister of the King of Prussia in London.

[p] It can be found in Collins's *Commercium Epistolicum*, published in London in 1712.

[q] He was the father of the current Lord Radnor. As early as 1693, he had informed the Royal Society of several problems concerning lotteries. Twenty years before Mr. De Montmort's essay was published, he had drawn up a table for use in *the game of the three raffles*. [The game of Raffles is analyzed in all three editions of De Moivre's *Doctrine of Chances*; three Raffles uses three sets of three dice.]

[r] *Histoire de l'Académie des Scien*ces of 1719, page 89. However, it seems to me that at this point, Mr. De Fontenelle was speaking only of the initial impression that Mr. De Moivre's essay had made on Mr. De Montmort, and not the one that stayed with him after he read the second.

[s] *Lib.*, vol. VII.

[t] *Comm. Epist.*, vol, II, page 220.

[u] His discovery of these sequences follows closely on the heels of his essay on *The Measurement of Chance*; vol. XVIII. Some of their properties were inferred in the *Philosophical Transactions* of 1722, no 373, but their proofs can only be found in the *Miscellanea Analytica*.

[x] *Ars Conjectandi* Basel 1713 In Pt. 4.

[y] See his *Éloge* in the *Histoire de l'Académie des Scien*ces of 1705, page 149.

[z] Numerous trials had been made at Mr. De Moivre's request, and they confirmed his rule.

<sup>a</sup>As everything concerning great men may be of interest, it is perhaps worth noting that Mr. Newton often told Mr. De Moivre that if he had not been so old, he would have been tempted, in the light of his recent observations, *to have another pull at the moon* (i.e., to revise his theory of the moon). Mr. De Moivre himself related this to me. [The italicized phrase is in English in Maty's original.]

<sup>b</sup>Mr. Forney, Secretary of the Berlin Academy, kindly provided me with this information.

<sup>c</sup>See the *Philosophical Transactions*, nos 196 and 198.

<sup>d</sup>Could it not be conjectured that this increase, which takes place on a period of four to five years, is due to gender-related illnesses occurring during this critical period?

<sup>e</sup>Quoted verbatim from the preface to his book (De Moivre, 1743).

<sup>f</sup>Mr. Dodson. See his memoir in the *Philosophical Transactions* of 1752, vol. XLVII.

## REFERENCES

### Manuscript Sources

Bibliothèque nationale de France
  Procès verbaux de l'Académie royale des sciences, tome 73, 1754, pages 425 and 428
Carnegie Mellon University
  Posner Family Collection
Columbia University Library
  David Eugene Smith Collection: letter from Abraham de Moivre to Edward Montague
Centre for Kentish Studies
  U1590 C14/2. Correspondence with P. Murdoch
  U1590 C21. Papers by several eminent mathematicians addressed to or collected by Lord Stanhope (contains three letters from De Moivre to Stanhope)
  U1590 A98. Account of my Expenses beginning November 28th. 1735
Public Record Office
  C 104/266 Bundle 38: Papers of Daniel de Moivre relating to trade, mainly in precious stones and jewellery, in London and Vera Cruz, Mexico
  PROB 11/588: will of Peter Magneville
  PROB 11/661: will of Daniel De Moivre (senior)
  PROB 11/666: will of Daniel De Moivre (junior)
  PROB 11/811: will of Abraham De Moivre
Royal Society
  Classified Papers: Cl.P.I.43
  Election certificates: EC/1737/04 (George Lewis Scott); EC/1737/09 (John Peter Bernard); EC/1737/17 (Philip Naudé); EC/1738/11 (Hermann Bernard); EC/1739/10 (John Peter Stehelin); EC/1740/07 (Peter Davall); EC/1740/08 (Henry Stewart Stevens); EC/1741/03 (Francis Philip Duval); EC/1743/02 (Roger Paman); EC/1743/09 (Jean Masson); EC/1745/03 (Peter Wyche); EC/1745/09 (John Christopher Pepusch); EC/1745/18 (Edward Montagu); EC/1746/15 (Daniel Peter Layard); EC/1747/05 (David Ravaud); EC/1753/07 (Robert Symmer)
  Folkes Collection: ms. 250 letter from De Moivre to Folkes
  Journal Books of Scientific Meetings: Volumes 9, 10, 11
  Letter Book Volume 14: correspondence of John Shuttleworth
University of Chicago
  The Joseph Halle Schaffner Collection Box 1, Folder 51: Memorandum relating to Sir Isaac Newton given me by Mr. Abraham Demoivre in Novr. 1727
Westminster Council Archives
  Parish Register, St. Martin-in-the-Fields Church, 1754
  Poor Law Rate Books, St. Martin-in-the-Fields Parish, 1750–1754

### PRINTED SOURCES